\documentclass[useAMS,usenatbib]{mn2e}

\usepackage{epsfig}
\usepackage{graphicx}
\usepackage{psfig}
\usepackage{amsmath}

\newcommand{\lrmv}{\ell_{\rm rmv}}
\newcommand{\WMAP}{\textit{WMAP}}

\title[The Hot and Cold Spots in \WMAP5]
{The Hot and Cold Spots in Five-Year \WMAP\ Data}

\author[Hou, Banday, G\'orski]
{Zhen Hou$^{1,2,3}\footnote{E-mail:houzhen@mpa-garching.mpg.de}$,
A.J. Banday$^{2,4}$,
K.M. G\'orski$^{5,6,7}$\\
$^{1}$ Purple Mountain Observatory, Chinese Academy of Sciences, 210008, Nanjing, China\\
$^{2}$ Max-Planck-Institute for Astrophysics,
Karl-Schwarzschildstrasse 1, D-85748, Garching bei Muenchen, Germany\\
$^{3}$ Graduate University of Chinese Academy of Sciences, 100049,
Beijing, China\\
$^{4}$ Centre d'Etude Spatiale des Rayonnements, 9 av du Colonel Roche, BP 44346, 31028 Toulouse Cedex 4, France\\
$^{5}$ Jet Propulsion Laboratory, 4800 Oak Grove Drive, Pasadena CA 91109, USA\\
$^{6}$ California Institute of Technology, Pasadena, CA 91125, USA\\
$^{7}$Warsaw University Observatory, Aleje Ujazdowskie 4, 00-478 Warszawa, Poland\\
}

\pagerange{\pageref{firstpage}--\pageref{lastpage}}

\pubyear{2009}

\begin{document}

\maketitle

\label{firstpage}

\begin{abstract}
We present an extensive frequentist analysis of the one-point
statistics (number, mean, variance, skewness and kurtosis) and
two-point correlation functions determined for the local extrema of
the cosmic microwave background temperature field observed in
five-years of \textit{Wilkinson Microwave Anisotropy Probe} (\WMAP)
data. Application of a hypothesis test on the one-point statistics
indicates a low-variance of hot and cold spots in all frequency
bands of the \WMAP\ data. The consistency of the observations with
Gaussian simulations of the the best-fitting cosmological model is
rejected at the 95\% C.L. outside the \WMAP\ KQ75 mask and the
northern hemispheres in the Galactic and ecliptic coordinate frames.
We demonstrate that it is unlikely that residual Galactic foreground
emission contributes to the observed non-Gaussianities. However, the
application of a high-pass filter that removes large angular scale
power does improve the consistency with the best-fitting
cosmological model.

Two-point correlation functions of the local extrema are calculated
for both the temperature pair product (T-T) and spatial
pair-counting (P-P). The T-T observations demonstrate weak
correlation on scales below $20^\circ$ and lie completely below the
lower $3\sigma$ confidence region once various temperature
thresholds are applied to the extrema determined for the KQ75 mask
and northern sky partitions. The P-P correlation structure
corresponds to the clustering properties of the temperature extrema,
and provides evidence that it is the large angular-scale structures,
and some unusual properties thereof, that are intimately connected
to the properties of the hot and cold-spots observed in the \WMAP\
five-year data.

\end{abstract}

\footnotetext{E-mail: houzhen@mpa-garching.mpg.de}

\section{Introduction}
\label{intro}

Generic inflationary theories predict that the initial conditions of
the universe are Gaussian random fields with a nearly scale
invariant or Harrison-Zel'dovich spectrum. The cosmic microwave
background (CMB) carries the imprint of such random fields in the
temperature fluctuations observed from the last-scattering surface.
The statistical properties of the observed CMB sky, by comparison
with theoretical predictions, thus provide a mechanism to test
whether the early universe is consistent with the Gaussian
hypothesis or whether a primordial non-Gaussian component is
required.

In this paper, we consider the detailed properties of local extrema,
defined as those pixels whose temperature values are higher
(hotspots) or lower (coldspots) than all of the adjacent (neighbour)
pixels \citep{wandelt_etal_1999}. The connection between the maxima
of Gaussian random fields and CMB temperature fluctuations resulting
from gravitational fluctuations has long been established
\citep{zabotin_etal_1985, sazhin_1985a,sazhin_1985b}. A
comprehensive study of the statistical properties of local extrema
in the CMB temperature field, including the number density and
angular correlation function of the peaks (hotspots) and troughs
(coldspots) for Gaussian random processes was undertaken by
\citet{bond_etal_1987} and \citet{vittorio_etal_1987}.
\citet{coles_etal_1987} extended these predictions for various toy
(Rayleigh, Maxwell, Chi-squared, lognormal, rectangular and Gumbel
type I) non-Gaussian random fields including statistics such as the
mean size and frequency of occurrence of upcrossings, and discussed
whether it is possible to determine if the observed statistics of
the CMB sky are indeed Gaussian as predicted by standard
inflationary theory. \citet{barreiro_etal_1997} showed that the
number and Gaussian curvature of local extrema valid for a given
threshold were sensitive enough to distinguish the geometry of the
universe. The temperature-correlation function of CMB hotspots above
a certain threshold has been accurately calculated by
\citet{heavens_etal_1999} on small-angle separations in the
flat-plane approximation, and subsequently extended to full-sky
coverage in \citet{heavens_etal_2001}. It is considered that these
correlations provide a test of the Gaussian hypothesis of initial
conditions and can discriminate  between  inflation and topological
defect models.

This statistical analysis technique has been applied to three types of
observations by \citet{matinez_etal_1989} and to study the Tenerife
Experiment data along a strip in the CMB sky
\citep{gutierrez_etal_1994}. \citet{kogut_etal_1995,kogut_etal_1996}
used the properties of local extrema to test the consistency of the
\textit{COBE}-DMR
data to the predictions of inflationary cosmological models,
also including a comparison to several toy non-Gaussian models.

Data from the \textit{Wilkinson Microwave Anisotropy Probe} (\WMAP)
currently provide the most comprehensive, full-sky, high-resolution
information on the CMB. \citet{LW04} made the first analysis of the
statistical properties of hot and cold spots in the first-year of
\WMAP\ data computing their number, mean, variance, skewness and
kurtosis values.  Their main conclusions were that the mean excursion
of hot and cold spots were not hot and cold enough respectively
compared to their Gaussian simulations, and there was also evidence
for low variance in the northern ecliptic hemisphere.  In a subsequent
paper, they developed a hypothesis test schema in order to study the
robustness of the earlier claims, and particularly evaluated the
dependence of the earlier work to variations in the noise model
assumptions and to the resolution of the maps \citep{LW05}. In
addition, a $3\sigma$-level anomaly has been detected using the
temperature-correlation functions of the local
extrema. \citet{tojeiro_etal_2006} concentrated on the properties of
the extrema point-correlation function using a technique well-known in
galaxy clustering studies. Evidence for non-Gaussianity was found on
large-scales, but its origin was not definitely established.

In this paper, we analyse both the one- and two-point statistics of
local extrema in the five-year release of \WMAP\ data. The
hypothesis test introduced by \cite{LW05} is adopted in our
one-point analysis to make a statement at a certain significance
level regarding whether the observed statistics are consistent with
our Gaussian random simulations. Further analysis, using additional
equatorial Galactic cuts, and the removal of specific low-$\ell$
modes, has been carried out to elucidate the origin of these
anomalies. We undertake both temperature- (T-T) and point- (P-P)
correlation analyses to further confirm the findings of our
one-point analysis and relate the correlation structures to spatial
temperature distributions. This paper is organized as follows. In
Section \ref{the_maps} we present an overview of the \WMAP\ data
used in the analysis and key properties necessary to allow
the generation of Gaussian realizations having identical instrumental properties
as the data. Section \ref{the_masks} details the masks adopted in order to
minimise contamination from non-cosmological
sources, \ref{l_dependence} prescribes the technique used to
subtract large-angular scale structure to test the sensitivity of the
results to putative anomalous features therein,
and \ref{simulations} describes the process for simulating the CMB
sky in a manner consistent with the \WMAP\ data.  Section
\ref{maps_LE} specifies the method for determining local extrema
from the observed sky maps and simulations, and the statistics
used in our analysis are specified in section \ref{statistics}.
Results are reported in Section \ref{results},
including the analysis and discussion of one-point statistics (Section
\ref{1p_results}) and two-point correlation functions (Section
\ref{2p_results}).  Finally, we present our conclusions in Section
\ref{conclusion}.

\section{METHOD}
\label{get_LE}

Even though the literature contains extensive theoretical
predictions of the statistics of local extrema in the CMB, such as
the number density and two-point correlation functions, a
frequentist approach based on simulated statistics to be compared
with the corresponding values for the \WMAP\ data is indicated here,
since the inhomogeneous observing strategy and complicated
sky-coverage of the mask used during data analysis are difficult to
account for analytically. This section provides the key information
required for such an analysis and comparison.

\subsection{The WMAP data} \label{the_maps}

The \WMAP\ instrument is composed of 10 differencing assemblies
(DAs) spanning five frequencies from 23 to 94~GHz
\citep{bennett_etal_2003}. The two lowest frequencies (K and Ka) are
generally used as Galactic foreground monitoring bands, with the
three highest (Q, V, and W) being available for cosmological
assessment. Note, however, that the Q-band information is dropped by
the \WMAP\ team for their power-spectrum analysis of the 5-year
data. There are 2 high-frequency DAs for Q band (Q1, Q2), 2 for V
(V1, V2), and 4 for W (W1,W2,W3,W4), each with a corresponding beam
profile and characteristic noise properties.  The maps are provided
in the HEALPix pixelization scheme \citep{gorski_etal_2005}, with
resolution parameter $N_{\rm side}=512$. We utilize the 5-year
foreground-reduced temperature maps available from the LAMBDA
website\footnote{http://lambda.gsfc.nasa.gov/product/map/dr3/maps\_da\_forered\_r9\_iqu\_5yr\_get.cfm}.

The instrumental noise can be considered to consist mainly of two parts:
a white noise contribution, and a $1/f$ component.
The noise in the \WMAP\ sky maps
is weakly correlated as a consequence of the differential nature of
the observations, the inhomogeneous scanning strategy and the
$1/f$ term. This biases
measurements on certain low-$\ell$ modes, although it remains
unimportant for temperature analysis because of the high signal-to-noise
on these scales. The effect is not important at high-$\ell$
\citep{hinshaw_etal_2007}. Thus we consider that the noise can be
entirely described by an uncorrelated instrumental white noise
component with rms value per pixel given by
\begin{equation}
\sigma_{i}(\textbf{\textit{n}})=\frac{\sigma_{0,i}}{\sqrt{N_{{\rm
obs},i}(\textbf{\textit{n}})}}
\end{equation}
where $\sigma_{0,i}$ is the rms noise per observation for a given DA
(as tabulated in \citet{hinshaw_etal_2008}), and $N_{{\rm
obs},i}(\textbf{\textit{n}})$ is the number of observations at a
given pixel. The scan pattern is such that the latter is greatest at
the ecliptic poles and in rings surrounding them, and fewest in the
ecliptic plane \citep{bennett_etal_2003}.

For analysis purposes, we average the individual DAs at a given
frequency for the Q, V and W bands using uniform and equal weights
over all pixels rather than noise weighting (as utilized for some
aspects of the \WMAP\ power spectrum analysis), since this results
in a simple effective beam at each frequency. We also form simple
combinations of the least foreground-contaminated V and W bands
(VW), and all of the Q, V, W band data (QVW).

\subsection{Masks} \label{the_masks}

To avoid contamination of the cosmological signal by emission from
the diffuse Galactic foreground and distant point sources, the
\WMAP\ mask for extended temperature analysis (KQ75, roughly 72\%
sky coverage) is applied to the data. We also extend the KQ75 mask
to include data only from specific hemispheres -- Galactic North and
South (GN, GS), and Ecliptic North and South (EN, ES). In the
remainder of the paper, we will often refer to results derived on
the KQ75 sky coverage as \lq full-sky' results for convenience, and
to distinguish from those values computed with additional hemisphere
masking. The possibility that any evidence for non-Gaussianity can
be associated with residual Galactic foregrounds implies the need
for tests utilizing more aggressive pixel rejection. We construct
additional masks which exclude various symmetric latitude cuts,
specifically $|b|<15^{\circ}$, $20^{\circ}$, $25^{\circ}$,
$30^{\circ}$, in addition to the KQ75 regions to help constrain our
one-point statistics.

\subsection{$\ell$-dependence} \label{l_dependence}

As is now standard practise in CMB data analysis, we subtract the
best-fitting monopole and dipole terms from the data.

However, the low value of the quadrupole as observed in the \WMAP\
5-year data, together with the observed strong alignment between the
quadrupole and octopole moments (possibly extending to even higher
orders) motivates the removal of the quadrupole in addition to the
monopole and dipole.  As in previous studies by
\citet{tojeiro_etal_2006}, we also remove higher moments for some
specific comparisons, and in particular we consider the ranges
$\ell=$ 0--5 and $\ell=$ 0--10.

The fitting method is a $\chi^2$ minimization technique.  For
cut-sky fitting, the multipoles are coupled and there are
differences between the corresponding derived multipoles for
different fitting ranges. For example, we fit multipoles $\ell=$
0--2 and $\ell=$ 0--5, then the dipole or quadrupole realizations
fitted by these two are different. However, since our aim is not to
determine the actual values of these low order multipoles, and given
that the subtraction is applied consistently to both the \WMAP\ data
and each simulation, the differences should not be relevant or bias
the analysis.

\subsection{Simulations} \label{simulations}

We generate a large number of simulations that should represent
accurate approximations of the assumed Gaussian primordial
temperature fluctuations combined with the \WMAP\ observation
properties (beam resolution and noise amplitudes). These should
provide sufficient reference statistics to allow us to search for
evidence of primordial non-Gaussianity in the data.

We perform each Gaussian simulation to yield a map for further
analysis in the following way.

\begin{enumerate}
\renewcommand{\theenumi}{\arabic{enumi}}
\item We generate the array of spherical harmonic coefficients $a_{lm}$
as a set of Gaussian random numbers with variance defined by the
\WMAP5 best-fitting `lcdm+sz+lens' model power spectrum. Since we
wish to create maps with the \WMAP\ resolution level $N_{\rm
side}=512$, the set of coefficients is truncated corresponding to a
maximum multipole $\ell_{\rm max}=1024$. The corresponding
pixelization window function $p_l$ is applied to each $a_{lm}$
coefficient. Similarly, the appropriate beam transfer function for
each DA, $b_l$, is also applied to obtain the coefficients
$\tilde{a}_{lm}=b_{l}p_{l}a_{lm}$.

\item The CMB sky realization is created pixel-by-pixel by
\begin{equation}
T(\textbf{\textit{n}})=\sum_{l=0}^{l_{\rm
max}}\sum_{-m}^{m}\tilde{a}_{lm}Y_{lm}(\textbf{\textit{n}}).
\end{equation}

\item We create a noise realization for each DA by combining
a set of Gaussian random numbers, $g(\textbf{\textit{n}})$, with
zero mean and unit variance with the expected noise rms
pixel-by-pixel
$N(\textbf{\textit{n}})=g(\textbf{\textit{n}})\sigma_{b}(\textbf{\textit{n}})$,
then add it to the CMB realization $T(\textbf{\textit{n}})$.

\item We set the temperature of pixels
lying inside the masked regions defined previously to
the HEALPix sentinel value.
\end{enumerate}

Now, we have both the \WMAP\ measurements and simulated realizations
of them to analyse. The processing method hereafter is identical for
both of them.

\subsection{Analysis} \label{maps_LE}

In this section, we introduce the processing steps necessary for
allowing the study of the local extrema in either the data or
simulated maps as specified in Section~\ref{the_maps}. Generally,
this closely follows the scheme defined in Section 4 of
\citet{LW05}, which is in summary

\begin{enumerate}
\renewcommand{\theenumi}{\arabic{enumi}}
\item We fit and subtract the monopole and dipole contributions, together with
the quadrupole and higher multipoles ($\ell=$ 0--5 or $\ell=$ 0--10
for certain bands) if required, from maps outside the mask region.

\item We smooth the maps in each band,
as well as any mask to be applied, with a Gaussian beam of
FWHM as defined in Section~\ref{definefwhm} and recorded in Table \ref{fwhm_5yr}.

\item On the region of valid pixels (defined in
Section~\ref{validpixels}), we find the local extrema and compute
the temperature standard deviation of all the valid pixels,
$\sigma_{\rm sky}$.
\end{enumerate}

\begin{figure}
\includegraphics[width=0.48\textwidth,trim=0.5cm 0 0.8cm 1cm]{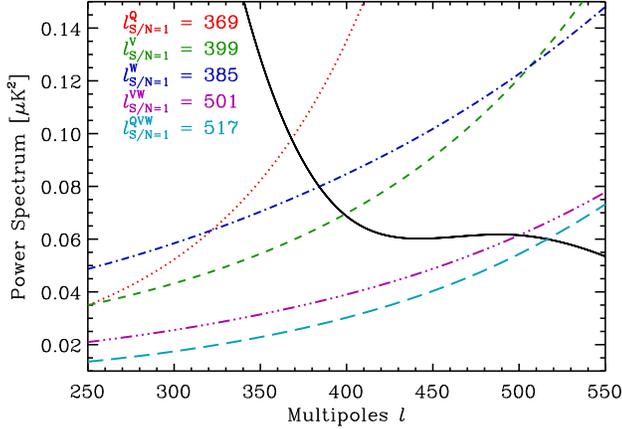}
\caption{Determination of smoothing FWHM. This is achieved by
evaluating the $\ell$-value satisfying the criterion that
signal-to-noise is unity ($\rm S/N=1$). The black-solid curve is the
\WMAP5 best-fitting `lcdm+sz+lens' model power spectrum excluding
any correction for pixelization and instrumental beam window
functions; the red-dotted, green-dashed, blue-dash-dot,
purple-dash-dot-dot, and lightblue-long-dashes curve are averaged
noise spectra for the Q, V, W, VW and QVW bands respectively,
deconvolved from pixelization and instrumental beams to be
consistent with the theoretical power spectrum. The $\ell_{\rm
S/N=1}$ values can be determined by the position of the crossing
points.} \label{SN}
\end{figure}

\subsubsection{Applied smoothing} \label{definefwhm}
Smoothing is applied in our analysis because it enhances the
signal-to-noise ratio (S/N) and removes the local sensitivity to the
fine-structure of the noise. The angular scale of smoothing is
determined by a S/N normalization criterion as shown in Figure
~\ref{SN}. For each band, the observed power spectrum,
$\tilde{C}_{l}$, is related to the underlying spectrum $C_{l}$, on
average, as
\begin{equation}
\langle\tilde{C}_{l}\rangle=b_{l}^{2}p_{l}^{2}C_{l}+\langle
N_{l}\rangle,
\end{equation}
where $N_{l}$ is the noise power spectrum of each band. We find the
$\ell$ values corresponding to ${\rm S/N}=1$, i.e.,
$C_{l}=b_{l}^{-2}p_{l}^{-2}\langle N_{l}\rangle$, for the Q, V, W, VW
and QVW bands, listed in Figure ~\ref{SN}. The smoothing suppression
factor in $\ell$-space is
\begin{equation}
g(\ell,\theta_{\rm FWHM})=\exp(-\frac{1}{2}\ell(\ell+1)\sigma^2),
\label{gauss_fwhm}
\end{equation}
where $\sigma=\theta_{\rm FWHM}/\sqrt{8\log2}$, and $\theta_{\rm FWHM}$
is the Gaussian FWHM scale in radians. We choose two different FWHM
scales ($\theta_{\rm f1}$ and $\theta_{\rm f2}$) for each band such that
$g(l_{\rm S/N=1}^{b},\theta_{\rm f1,f2})=e^{-1},10^{-1}$. The FWHM
values for the 5-year S/N normalization are listed in arcmin in Table
\ref{fwhm_5yr}.

\begin{table}
\caption{The Gaussian FWHM scale (arcmin) of all the bands for
5-year S/N normalization.}
\begin{center}
\begin{tabular}{r|ccccc}
\hline
& Q & V & W & VW & QVW \\
\hline
$\theta_{\rm f1}$ & 30.984 & 28.657 & 29.698 & 22.828 & 22.123 \\
$\theta_{\rm f2}$ & 47.015 & 43.485 & 45.064 & 34.640 & 33.569 \\
\hline
\end{tabular}
\end{center}
\label{fwhm_5yr}
\end{table}

\subsubsection{Valid pixels} \label{validpixels}

After smoothing, we are concerned that edge effects due to pixels
close to the boundaries of the mask may create artefacts in the
local extrema measured in these regions. To solve this problem, we
apply the method used by \citet{eriksen_etal_2005}. The original
mask is convolved with a Gaussian FWHM beam as in
Eq.\ref{gauss_fwhm} and valid pixels for further analysis are
defined as those with a value larger than 0.90 in the smoothed mask.
We have tested that 0.90 is sufficiently conservative in our local
extrema analysis.

\subsection{The statistics} \label{statistics}
In this section, we introduce the statistics by which a comparison
will be made between the \WMAP\ 5-year data and simulations, as well as the
methodology applied in order to establish confidence levels on
these statistics.

\subsubsection{One-point statistics} \label{1p_sta}
Following the analysis performed by \citet{LW04}, we calculate the
number, mean, variance, skewness and kurtosis value of local
extrema. The mean and skewness values of the cold spots are
multiplied by $-1$ for convenience when comparing with hotspots. We
compute the fraction of simulated statistical values that are lower
than those for the \WMAP\ data and use these numbers to make a
statement about whether the measurements are consistent with the
Gaussian process assumed for the simulations.

To achieve this, we adopt the hypothesis test methodology introduced
by \citet{LW05}, and extensively described in their appendix 2, to
establish statistically robust confidence limits for this assessment
of our one-point analysis.

As a brief summary of this methodology, consider that we have one
number, $x_{0}$, the mean value determined from the observed local
extrema, and a sample, $\{x_{i}\}$, from the simulations. We
consider the hypothesis `$x_{0}$ comes from the same distribution as
$\{x_{i}\}$'. If there are $n$ total samples in the simulated
ensemble, and $i$ do fall below $x_{0}$, then we can define the
quantity $p$ as $i/n$ where $p$ defines the position of $x_{0}$ in
the appropriate PDF. We then choose a hypothesis about $p$
\begin{equation}
H:p\in(\alpha/2,1-\alpha/2),
\end{equation}
where $\alpha$ is the significance level much less than 1. It is a
two-sided hypothesis because $x_{0}$ is less likely to come from the
distribution than $\{x_{i}\}$ if it lies on the tail of $\{x_{i}\}$
or beyond, as naturally considered. The frequentist statement is
that `$H$ is true in a fraction $1-\alpha$ of all possible Gaussian
universes'. We make tests for each statistic twice: first, the Type
\uppercase\expandafter{\romannumeral1} error (rejection of true
hypothesis) probability is controlled to be small; and second, the
Type \uppercase\expandafter{\romannumeral2} error (acceptance of the
false hypothesis) is controlled. See Appendix 2 of \citet{LW05} for
comprehensive details.

\subsubsection{Two-point statistics} \label{2p_sta}

There are two kinds of two-point analysis performed in this paper:
the temperature- (T-T) and point- (P-P) angular correlation functions.

The T-T correlation function discussed here has the same definition
as that discussed by
\citet{eriksen_etal_2005},
\begin{equation}
\xi_{\rm TT}(\theta)=\langle
T(\textbf{\textit{n}}_1)T(\textbf{\textit{n}}_2)\rangle
\end{equation}
where $\cos\theta=\textbf{\textit{n}}_1\cdot\textbf{\textit{n}}_2$,
and the points  $\textbf{\textit{n}}_1$ and $\textbf{\textit{n}}_2$
at which the temperatures are defined correspond to local extrema.
We analyse the maxima and minima separately, i.e., only the
maxima-maxima ($T_{\rm max}-T_{\rm max}$) and minima-minima ($T_{\rm
min}-T_{\rm min}$) correlation functions are evaluated.

The P-P angular correlation function is defined as the excess
probability of finding a pair of local extrema at angular separation
$\theta$ by direct analogy with the definition used in galaxy
distribution studies,
\begin{equation}
dP=\bar{n}[1+\xi_{\rm PP}(\theta)]d\Omega,
\end{equation}
where $\bar{n}$ is the mean number density of our sample. The
Hamilton estimator of the P-P correlation function is adopted here
\citep{ham_1993},
\begin{equation}
\xi_{\rm PP}(\theta)=\frac{DD(\theta)RR(\theta)}{[DR(\theta)]^2}-1,
\label{ham_est}
\end{equation}
where $DD(\theta)$ is the number of pairs of local extrema in our
data (for either the \WMAP\ observations or a particular simulation)
inside the interval $[\theta,\theta+d\theta)$, $RR(\theta)$ is the
number of pairs in a random generated sample with separation in the
same interval, and $DR(\theta)$ has the same meaning but the pairs
are selected between the data and the random sample.

We generate 200000 random points uniformly distributed on the full
sphere. We eliminate any adjacent points since these cannot
correspond to local extrema as defined in this work. This leaves
135889 surviving points, still at least ten times more numerous than
the average number in our data catalog, which is  usually considered
sufficient for the correlation function study.

$\chi^2$ values are computed for both T-T and P-P correlation
functions to quantify the degree of agreement between observations
and simulations. The $\chi^2$ statistic is calculated from the
difference between each sample (either observation or simulation)
and the average of the simulations on a given scale and defined as
\begin{equation}
\chi^{2}=\sum^{N_{\rm bin}}_{i,j=1}[\xi(i)-\langle
\xi(i)\rangle]M^{-1}_{ij}[\xi(j)-\langle \xi(j)\rangle],
\label{eq_chisqr}
\end{equation}
where $M^{-1}_{ij}$ is the inverse covariance matrix estimated from
the ensemble of correlation functions determined from the Gaussian simulations.
\begin{equation}
M_{ij}=\frac{1}{N_{\rm sim}-1}\sum_{n=1}^{N_{\rm
sim}}[\xi_{n}(i)-\langle \xi(i)\rangle][\xi_{n}(j)-\langle
\xi(j)\rangle].
\end{equation}
where $N_{\rm sim}$ is the total number of simulations.

The $\chi^2$ statistic is optimized for studies on CMB $N$-point
(especially even-ordered) correlation functions
\citep{eriksen_etal_2004, eriksen_etal_2005} since they are strongly
asymmetrically distributed. Each 2-point configuration of the full
correlation function is
transformed by the relation
\begin{equation}
\frac{\rm Rank \ of \ observed \ map}{\rm{Total \ number \ of \
maps}+1}=\frac{1}{\sqrt{2\pi}}\int_{-\infty}^{s}e^{-\frac{1}{2}t^2}dt
\end{equation}
The numerator of the left hand side is the number of realizations with a
lower value than the current one, and the denominator is the total
number of realizations plus 1 in order to make $s$ symmetrically
distributed around 0 and to avoid an infinite confidence assignment for
the largest value. The $\chi^2$ value and the covariance matrix are
computed from the transformed configurations of the correlation function.

\section{RESULTS AND DISCUSSIONS} \label{results}

\subsection{One-point results} \label{1p_results}

The one-point statistics establish the shape of the temperature
distribution function for local extrema. For example, skewness
represents the degree of symmetry between the left and right tails
of the probability distribution, and kurtosis measures its
`peakedness'.

\subsubsection{General results} \label{general_1p_results}
Thousands of simulations are performed for five bands on two
smoothing scales with the low-$\ell$ multipoles removed in the range
$[0,\lrmv]$, with $\lrmv =1,2$. Five kinds of one-point statistic --
the number, mean, variance, skewness and kurtosis -- of local maxima
and minima, are calculated. The frequencies of the simulated
statistics lower than the observed values are shown in Table
\ref{1p_tab_nm} and Table \ref{1p_tab_vk}. As mentioned in Section
\ref{1p_sta}, a hypothesis test is performed on each frequency
twice, with significance level $\alpha=0.05$. First, the probability
of a Type \uppercase\expandafter{\romannumeral1} error is controlled
to be as small as $\alpha$, and the frequencies rejected by our test
are marked by an asterisk; secondly, we control the probability of a
Type \uppercase\expandafter{\romannumeral2} error to be as small as
$\alpha$, then the frequencies rejected this time but accepted by
the first test are marked by a question mark. For the cases marked
by an asterisk, we assert that the data is not consistent with our
Gaussian process, with an associated probability $5\%$ of a Type
\uppercase\expandafter{\romannumeral1} error.

We ignore the skewness results because no rejection occurs. Tests
reject several cases for the number of local extrema, most of which
occur on smaller smoothing scales ($\theta_{\rm f1}$) and in the
northern hemisphere for cold spots. The mean values are quite
consistent with simulations at the $95\%$ confidence level, except
for rejections of the W band with $\theta_{\rm f2}$ smoothing in the
Galactic northern hemisphere, implying that the cold spots are not
cold enough. Other bands with $\theta_{\rm f2}$ smoothing show a
similar suppression of cold spots, although the values are accepted
by our test.

The main one-point abnormality appears in the variance results.
There is also significant hemispherical asymmetry indicated in Table
\ref{1p_tab_vk} -- almost all of the full-sky (NS) and northern (GN
and EN) results are strongly rejected for $\lrmv=1$, as well as GN
and most of EN for $\lrmv=2$. The frequencies with values outside
the $3\sigma$ confidence region have been detected and underlined.
It is worth noting that the measured variance on the full-sky is
more consistent with simulations after quadrupole removal, and also
slightly improved for the  northern Ecliptic hemisphere, which give
us evidence that the observed low-quadrupole is a possible source of
such anomalous behaviour, but this may not be the complete
explanation.

\subsubsection{Residual Galactic foreground}

The conventional wisdom suggests that evidence for non-Gaussianity
may be associated with residual Galactic foregrounds. We test this
simply by imposing additional masks with $|b|<15^{\circ}$,
$20^{\circ}$, $25^{\circ}$, $30^{\circ}$ symmetric cuts around the
Galactic equator to constrain the one-point statistics. We choose
the V band and smoothing scale $\theta_{\rm f2}$ to carry out the
test. The maxima and minima in the Galactic-cut region will then not
be involved in the conclusions. Table \ref{tab_galcut_rl0102}
presents the statistics after imposing the Galactic cuts, with
monopole and dipole, as well as quadrupole subtraction. Since the
skewness is unrevealing, it will not be listed again.

The results are generally similar to the previous cases. In
particular, the variance remains as asymmetric as ever. Although in
some cases the low-variance improves with increasing Galactic cut,
it is difficult to assert whether this is significant. Therefore, it
seems unlikely that residual Galactic foregrounds constitute a
solution to the anomalies seen in the one-point statistics.

\subsubsection{Low-$\ell$ subtraction}

Motivated by the quadrupole evidence in Section
\ref{general_1p_results}, we proceed to remove higher moments of
$\ell=$ 0--5 and $\ell=$ 0--10 separately from observed and
simulated maps on V band. The number, mean, variance and kurtosis of
hotspots and cold spots for $\lrmv=5$ and $\lrmv=10$ are put in
Table \ref{1p_tab_nm} and Table \ref{1p_tab_vk}.

The number, mean and variance show good agreement with simulations
for $\lrmv=5$ and even better for $\lrmv=10$, but problems of the
kurtosis for $\lrmv=5$ show that the peak of the extrema temperature
distribution on ES is much sharper than the Gaussian one, and it
disappears after subtraction of the higher moments. This result
suggests that the anomalies of the extrema temperature distribution
are related to the large-scale moments, especially the first five.
However, it is difficult to further confirm how the large-scale
moments affect the temperature distribution of local extrema if only
the one-point analysis is available. A two-point analysis is
necessary, for which we will also investigate the properties of the
two-point correlation as a function of different temperature
intervals.

\begin{table*}
\caption{Frequencies of the extrema one-point statistics derived from simulations with
lower values than the \WMAP5\ data. NS, GN, GS, EN and ES
correspond to full-sky, Galactic North, Galactic South, Ecliptic
North and Ecliptic South sky-coverage outside KQ75 mask,
respectively. $\theta_{\rm f1,f2}$ gives the smoothing scale
described by step 2 in Section \ref{maps_LE}. The number 1000, 2000
or 10000 corresponds to how many simulations were performed. The values rejected
by the hypothesis test are marked by a * or ?. Frequencies outside the
$3\sigma$ confidence region are underlined.}
\begin{center}
\begin{tabular}{r|lllll|lllll}
 \hline
 \hline
 \multicolumn{1}{c}{NUMBER}
 & \multicolumn{5}{l}{$\lrmv=1$}
 & \multicolumn{5}{l}{$\lrmv=2$}\\

 \multicolumn{1}{c}{5yr, KQ75 MASK}
 & NS & GN & GS & EN & ES
 & NS & GN & GS & EN & ES\\
 \hline

Q, max ($\theta_{\rm f1}$,1000) & 0.980? & 0.894  & 0.953 & 0.945  & 0.943  & 0.989* & 0.852  & 0.974? & 0.937  & 0.965 \\
Q, min ($\theta_{\rm f1}$,1000) & 0.996* & 0.989* & 0.934 & 0.983* & 0.977? & 0.993* & 0.986* & 0.909  & 0.973? & 0.991* \\
W, max ($\theta_{\rm f1}$,1000) & 0.714  & 0.577  & 0.691 & 0.528  & 0.641  & 0.690  & 0.497  & 0.789  & 0.606  & 0.578 \\
W, min ($\theta_{\rm f1}$,1000) & 0.957  & 0.988* & 0.377 & 0.984* & 0.633  & 0.938  & 0.990* & 0.365  & 0.986* & 0.573 \\
QVW, max ($\theta_{\rm f1}$,1000) & 0.922 & 0.904 & 0.756 & 0.924 & 0.679  & 0.948  & 0.793  & 0.844  & 0.932  & 0.671 \\
QVW, min ($\theta_{\rm f1}$,1000) & 0.894 & 0.849 & 0.723 & 0.894 & 0.588  & 0.847  & 0.889  & 0.678  & 0.905  & 0.591 \\
V, max  ($\theta_{\rm f1}$,2000) & 0.8410 & 0.4940 & 0.9615 & 0.8630
& 0.6435 & 0.8880 & 0.2560 & 0.9695?
& 0.9045 & 0.6700 \\
V, min  ($\theta_{\rm f1}$,2000) & 0.7855 & 0.8390 & 0.5585 & 0.8310
& 0.6270 & 0.7535 & 0.9180 & 0.3680
& 0.8110 & 0.7905 \\
VW, max  ($\theta_{\rm f1}$,2000) & 0.8995 & 0.9065 & 0.6930 &
0.8980 & 0.6670 & 0.9365 & 0.8355 & 0.8300
& 0.9085 & 0.6490 \\
VW, min  ($\theta_{\rm f1}$,2000) & 0.7715 & 0.9735? & 0.1860 &
0.9225 & 0.4450 & 0.7490 & 0.9905*
& 0.1580 & 0.9010 & 0.3775 \\

Q, max ($\theta_{\rm f2}$,1000) & 0.889 & 0.744 & 0.871 & 0.959 &
0.673 & 0.859 & 0.654 & 0.811 & 0.941
& 0.600 \\
Q, min ($\theta_{\rm f2}$,1000) & 0.908 & 0.886 & 0.815 & 0.880 &
0.839 & 0.896 & 0.869 & 0.767 & 0.733
& 0.855 \\
W, max ($\theta_{\rm f2}$,1000)  & 0.769 & 0.468 & 0.872 & 0.824 &
0.679 & 0.824 & 0.583 & 0.883 & 0.852
& 0.656 \\
W, min ($\theta_{\rm f2}$,1000)  & 0.980? & 0.996* & 0.371 & 0.973?
& 0.812 & 0.966 & 0.995* & 0.333
& 0.944 & 0.858 \\
QVW, max ($\theta_{\rm f2}$,1000) & 0.766 & 0.625 & 0.685 & 0.788 &
0.610 & 0.839 & 0.410 & 0.772
& 0.814 & 0.675 \\
QVW, min ($\theta_{\rm f2}$,1000) & 0.881 & 0.819 & 0.769 & 0.917 &
0.676 & 0.821 & 0.794 & 0.784
& 0.880 & 0.803 \\
V, max ($\theta_{\rm f2}$,10000)  & 0.5574 & 0.2597 & 0.8034 &
0.5319 & 0.6107 & 0.6076 &
0.2190 & 0.8228 & 0.4592 & 0.6013 \\
V, min ($\theta_{\rm f2}$,10000)  & 0.9666 & 0.9618 & 0.7395 &
0.9721 & 0.9073 & 0.9704 &
0.9752? & 0.6951 & 0.9657 & 0.9070 \\
VW, max ($\theta_{\rm f2}$,2000)  & 0.3810 & 0.2230 & 0.5665 &
0.5890 & 0.3095 & 0.3235 & 0.1395 & 0.6250
& 0.5850 & 0.2025 \\
VW, min ($\theta_{\rm f2}$,2000)  & 0.7055 & 0.8150 & 0.4390 &
0.8625 & 0.5105 & 0.6785 & 0.8665 & 0.2880
& 0.8255 & 0.6595 \\

\hline
 & \multicolumn{5}{l}{$\lrmv=5$}
 & \multicolumn{5}{l}{$\lrmv=10$}\\

V, max ($\theta_{\rm f2}$,10000) & 0.6932 & 0.1272 & 0.9408 & 0.5051
& 0.6778 & 0.6951 & 0.2259
& 0.9135 & 0.3168 & 0.8733 \\
V, min ($\theta_{\rm f2}$,10000) & 0.9001 & 0.9419 & 0.5887 & 0.9264
& 0.5525 & 0.8438 & 0.9132
& 0.6724 & 0.8914 & 0.7488 \\
\hline
\hline
 \multicolumn{1}{c}{MEAN}
 & \multicolumn{5}{l}{$\lrmv=1$}
 & \multicolumn{5}{l}{$\lrmv=2$}\\

 \hline

Q, max ($\theta_{\rm f1}$,1000) & 0.332 & 0.249 & 0.461 & 0.498 &
0.359 & 0.387 & 0.414 & 0.557 & 0.538
& 0.252 \\

Q, min ($\theta_{\rm f1}$,1000) & 0.132 & 0.082 & 0.500 & 0.256 &
0.077 & 0.167 & 0.044 & 0.422 & 0.268
& 0.064 \\

W, max  ($\theta_{\rm f1}$,1000) & 0.486 & 0.237 & 0.689 & 0.610 &
0.515 & 0.568 & 0.488 & 0.785 & 0.688
& 0.478 \\

W, min  ($\theta_{\rm f1}$,1000) & 0.411 & 0.283 & 0.728 & 0.229 &
0.537 & 0.476 & 0.196 & 0.584 & 0.247
& 0.649 \\

QVW, max  ($\theta_{\rm f1}$,1000) & 0.178 & 0.057 & 0.529 & 0.263 &
0.236 & 0.161 & 0.132 & 0.631
& 0.381 & 0.261 \\

QVW, min  ($\theta_{\rm f1}$,1000) & 0.503 & 0.496 & 0.538 & 0.247 &
0.694 & 0.588 & 0.509 & 0.462
& 0.286 & 0.775 \\

V, max  ($\theta_{\rm f1}$,2000) & 0.3925 & 0.3290 & 0.4960 & 0.4475
& 0.4895 & 0.4295 & 0.4230 & 0.5035
& 0.4065 & 0.3590 \\

V, min  ($\theta_{\rm f1}$,2000) & 0.5445 & 0.4555 & 0.6845 & 0.5205
& 0.4105 & 0.6670 & 0.3950 & 0.7440
& 0.6025 & 0.4270 \\

VW, max  ($\theta_{\rm f1}$,2000) & 0.4230 & 0.2385 & 0.6145 &
0.3715 & 0.5690 & 0.4385 & 0.3130 & 0.7485
& 0.4270 & 0.5570 \\

VW, min  ($\theta_{\rm f1}$,2000) & 0.3995 & 0.2790 & 0.6950 &
0.1160 & 0.7160 & 0.4540 & 0.2660 & 0.6890
& 0.2130 & 0.7570 \\

Q, max ($\theta_{\rm f2}$,1000) & 0.186 & 0.203 & 0.310 & 0.137 &
0.322 & 0.170 & 0.500 & 0.234 & 0.258
& 0.357 \\

Q, min ($\theta_{\rm f2}$,1000) & 0.096 & 0.095 & 0.326 & 0.206 &
0.082 & 0.139 & 0.081 & 0.484 & 0.208
& 0.133 \\

W, max ($\theta_{\rm f2}$,1000) & 0.190 & 0.208 & 0.376 & 0.344 &
0.222 & 0.175 & 0.478 & 0.351 & 0.424
& 0.209 \\

W, min ($\theta_{\rm f2}$,1000) & 0.065 & 0.014* & 0.441 & 0.055 &
0.140 & 0.082 & 0.006* & 0.493
& 0.049 & 0.234 \\

QVW, max ($\theta_{\rm f2}$,1000) & 0.309 & 0.111 & 0.624 & 0.305 &
0.517 & 0.338 & 0.366 & 0.624
& 0.392 & 0.523 \\

QVW, min ($\theta_{\rm f2}$,1000) & 0.161 & 0.101 & 0.407 & 0.167 &
0.304 & 0.260 & 0.108 & 0.530
& 0.278 & 0.364 \\

V, max ($\theta_{\rm f2}$,10000) & 0.1653 & 0.1369 & 0.3753 & 0.1999
&
0.3296 & 0.1638 & 0.4857 & 0.4187 & 0.3162 & 0.2739 \\

V, min ($\theta_{\rm f2}$,10000) & 0.1066 & 0.0769 & 0.3814 & 0.1181
&
0.1713 & 0.1509 & 0.0425 & 0.4078 & 0.1352 & 0.1719 \\

VW, max ($\theta_{\rm f2}$,2000) & 0.2980 & 0.1280 & 0.7070 & 0.4125
& 0.4245 & 0.3790 & 0.2320 & 0.7705
& 0.5585 & 0.3730 \\

VW, min ($\theta_{\rm f2}$,2000) & 0.2790 & 0.0880 & 0.6940 & 0.2050
& 0.3485 & 0.3215 & 0.0475 & 0.7790
& 0.2455 & 0.5065 \\

\hline
 & \multicolumn{5}{l}{$\lrmv=5$}
 & \multicolumn{5}{l}{$\lrmv=10$}\\

V, max ($\theta_{\rm f2}$,10000) & 0.4486 & 0.6654 & 0.2473 & 0.3489
& 0.2955 & 0.4703 & 0.4729
& 0.3356 & 0.6542 & 0.2627 \\

V, min ($\theta_{\rm f2}$,10000) & 0.0720 & 0.0339 & 0.3632 & 0.1334
& 0.1788 & 0.1100 & 0.1764
& 0.4261 & 0.1804 & 0.1273 \\

\hline
\hline
\end{tabular}
\end{center}
\label{1p_tab_nm}
\end{table*}

\begin{table*}
\caption{Continued from Table \ref{1p_tab_nm}}
\begin{center}
\begin{tabular}{r|lllll|lllll}
 \hline
 \hline
 \multicolumn{1}{c}{VARIANCE}
 & \multicolumn{5}{l}{$\lrmv=1$}
 & \multicolumn{5}{l}{$\lrmv=2$}\\

 \multicolumn{1}{c}{5yr, KQ75 MASK}
 & NS & GN & GS & EN & ES
 & NS & GN & GS & EN & ES\\
 \hline

Q, max ($\theta_{\rm f1}$,1000) & 0.006* & 0.013* & 0.382 & 0.006* &
0.248 & 0.063 & 0.005* & 0.683
& 0.017* & 0.541 \\

Q, min ($\theta_{\rm f1}$,1000) & 0.006* & 0.002* & 0.482 & 0.005* &
0.325 & 0.077 & 0.004* & 0.851
& 0.030? & 0.629 \\

W, max  ($\theta_{\rm f1}$,1000) & 0.002* & 0.011* & 0.233 & 0.003*
& 0.187 & 0.018? & 0.006* & 0.507
& 0.007* & 0.444 \\

W, min  ($\theta_{\rm f1}$,1000) & \underline{0.001}* &
\underline{0.001}* & 0.433 & 0.003* & 0.226 & 0.027? &
\underline{0.000}* & 0.773
& 0.016* & 0.456 \\

QVW, max  ($\theta_{\rm f1}$,1000) & 0.007* & 0.018? & 0.339 &
0.006* & 0.276 & 0.077 & 0.018? & 0.667
& 0.024? & 0.572 \\

QVW, min  ($\theta_{\rm f1}$,1000) & 0.002* & \underline{0.001}* &
0.474 & 0.003* & 0.238 & 0.027? & \underline{0.000}* & 0.790
& 0.013* & 0.524 \\

V, max  ($\theta_{\rm f1}$,2000) & 0.0030* & 0.0090* & 0.3035 &
0.0035* & 0.2170 & 0.0340 & 0.0075*
& 0.6415 & 0.0110* & 0.5200 \\

V, min  ($\theta_{\rm f1}$,2000) & 0.0045* & 0.0015* & 0.4265 &
0.0025* & 0.2335 & 0.0520 & \underline{0.0010}*
& 0.7495 & 0.0235? & 0.5815 \\

VW, max  ($\theta_{\rm f1}$,2000) & \underline{0.0010}* & 0.0060* &
0.2445 & 0.0015* & 0.1795 & 0.0220? & 0.0035*
& 0.5360 & 0.0060* & 0.4250 \\

VW, min  ($\theta_{\rm f1}$,2000) & 0.0025* & \underline{0.0010}* &
0.3660 & 0.0040* & 0.1715 & 0.0320 & \underline{0.0000}*
& 0.6830 & 0.0160* & 0.4530 \\

Q, max ($\theta_{\rm f2}$,1000) & 0.009* & 0.004* & 0.494 & 0.003* &
0.359 & 0.082 & 0.005* & 0.813
& 0.011* & 0.688 \\

Q, min ($\theta_{\rm f2}$,1000) & 0.006* & 0.003* & 0.478 & 0.003* &
0.337 & 0.053 & \underline{0.001}* & 0.807
& 0.021? & 0.613 \\

W, max ($\theta_{\rm f2}$,1000) & 0.005* & 0.007* & 0.353 & 0.003* &
0.308 & 0.050 & 0.003* & 0.662
& 0.010* & 0.591 \\

W, min ($\theta_{\rm f2}$,1000) & 0.007* & 0.002* & 0.465 & 0.004* &
0.300 & 0.048 & \underline{0.001}* & 0.810
& 0.022? & 0.566 \\

QVW, max ($\theta_{\rm f2}$,1000) & 0.003* & 0.006* & 0.358 & 0.003*
& 0.227 & 0.021? & 0.004* & 0.644
& 0.011* & 0.515 \\

QVW, min ($\theta_{\rm f2}$,1000) & 0.005* & \underline{0.001}* &
0.510 & 0.002* & 0.338 & 0.057 & \underline{0.000}* & 0.846
& 0.018? & 0.648 \\

V, max ($\theta_{\rm f2}$,10000) & 0.0040* & 0.0021* & 0.4491 &
\underline{0.0009}*
& 0.3188 & 0.0564 & 0.0031* & 0.7875 & 0.0059* & 0.6949 \\

V, min ($\theta_{\rm f2}$,10000) & 0.0028* & \underline{0.0011}* &
0.4345 & \underline{0.0009}* &
0.2750 & 0.0343 & \underline{0.0013}* & 0.8188 & 0.0061* & 0.5911 \\

VW, max ($\theta_{\rm f2}$,2000) & 0.0055* & 0.0105* & 0.3425 &
0.0035* & 0.2425 & 0.0465 & 0.0090*
& 0.5935 & 0.0080* & 0.5040 \\

VW, min ($\theta_{\rm f2}$,2000) & 0.0065* & 0.0025* & 0.4445 &
0.0015* & 0.3110 & 0.0630 & 0.0025*
& 0.7925 & 0.0155* & 0.6145 \\

\hline
 & \multicolumn{5}{l}{$\lrmv=5$}
 & \multicolumn{5}{l}{$\lrmv=10$}\\

V, max ($\theta_{\rm f2}$,10000) & 0.0808 & 0.0716 & 0.6817 & 0.0443
& 0.4864
& 0.0953 & 0.2413 & 0.2928 & 0.2689 & 0.3853 \\

V, min ($\theta_{\rm f2}$,10000) & 0.1213 & 0.0342 & 0.6285 &
0.0229? & 0.4688
& 0.0463 & 0.1644 & 0.2450 & 0.1584 & 0.3040 \\

\hline
 \hline
 \multicolumn{1}{c}{KURTOSIS}
 & \multicolumn{5}{l}{$\lrmv=1$}
 & \multicolumn{5}{l}{$\lrmv=2$}\\

 \hline

Q, max ($\theta_{\rm f1}$,1000) & 0.247 & 0.373 & 0.104 & 0.195 &
0.532 & 0.231 & 0.259 & 0.327 & 0.082
& 0.648 \\

Q, min ($\theta_{\rm f1}$,1000) & 0.515 & 0.140 & 0.575 & 0.290 &
0.824 & 0.752 & 0.062 & 0.884 & 0.309
& 0.820 \\

W, max  ($\theta_{\rm f1}$,1000) & 0.567 & 0.755 & 0.254 & 0.642 &
0.608 & 0.643 & 0.596 & 0.479 & 0.559
& 0.724 \\

W, min  ($\theta_{\rm f1}$,1000) & 0.716 & 0.613 & 0.465 & 0.493 &
0.804 & 0.800 & 0.382 & 0.675 & 0.493
& 0.847 \\

QVW, max  ($\theta_{\rm f1}$,1000) & 0.645 & 0.861 & 0.221 & 0.756 &
0.621 & 0.741 & 0.732 & 0.660
& 0.715 & 0.821 \\

QVW, min  ($\theta_{\rm f1}$,1000) & 0.735 & 0.579 & 0.560 & 0.571 &
0.791 & 0.818 & 0.234 & 0.756
& 0.435 & 0.913 \\

V, max  ($\theta_{\rm f1}$,2000) & 0.7010 & 0.9090 & 0.2355 & 0.7250
& 0.6925 & 0.8130 & 0.8690 & 0.6605
& 0.7105 & 0.8420 \\

V, min  ($\theta_{\rm f1}$,2000) & 0.7055 & 0.6930 & 0.4000 & 0.6395
& 0.7935 & 0.8015 & 0.4760 & 0.6505
& 0.6335 & 0.8845 \\

VW, max  ($\theta_{\rm f1}$,2000) & 0.4770 & 0.7145 & 0.1925 &
0.6290 & 0.5885 & 0.6125 & 0.5700 & 0.6410
& 0.6430 & 0.7670 \\

VW, min  ($\theta_{\rm f1}$,2000) & 0.6050 & 0.6375 & 0.3105 &
0.5475 & 0.6420 & 0.6560 & 0.2700 & 0.4720
& 0.4250 & 0.7020 \\

Q, max ($\theta_{\rm f2}$,1000) & 0.742 & 0.826 & 0.382 & 0.824 &
0.693 & 0.801 & 0.719 & 0.642 & 0.781
& 0.846 \\

Q, min ($\theta_{\rm f2}$,1000) & 0.440 & 0.700 & 0.085 & 0.482 &
0.598 & 0.682 & 0.363 & 0.437 & 0.520
& 0.756 \\

W, max ($\theta_{\rm f2}$,1000) & 0.822 & 0.850 & 0.604 & 0.897 &
0.784 & 0.890 & 0.715 & 0.733 & 0.881
& 0.845 \\

W, min ($\theta_{\rm f2}$,1000) & 0.881 & 0.942 & 0.381 & 0.579 &
0.932 & 0.932 & 0.774 & 0.642 & 0.615
& 0.942 \\

QVW, max ($\theta_{\rm f2}$,1000) & 0.477 & 0.483 & 0.264 & 0.536 &
0.533 & 0.514 & 0.390 & 0.569
& 0.454 & 0.668 \\

QVW, min ($\theta_{\rm f2}$,1000) & 0.705 & 0.584 & 0.424 & 0.514 &
0.811 & 0.800 & 0.304 & 0.646
& 0.483 & 0.911 \\

V, max ($\theta_{\rm f2}$,10000) & 0.8272 & 0.8613 & 0.5090 & 0.7485
&
0.8380 & 0.8448 & 0.7720 & 0.6555 & 0.6487 & 0.8630 \\

V, min ($\theta_{\rm f2}$,10000) & 0.8999 & 0.9194 & 0.5643 & 0.8362
&
0.8908 & 0.9741? & 0.8131 & 0.8130 & 0.8565 & 0.9359 \\

VW, max ($\theta_{\rm f2}$,2000) & 0.5245 & 0.6395 & 0.2705 & 0.4855
& 0.6195 & 0.5710 & 0.4900 & 0.6170
& 0.4370 & 0.8310 \\

VW, min ($\theta_{\rm f2}$,2000) & 0.7000 & 0.6475 & 0.4175 & 0.4475
& 0.8770 & 0.8170 & 0.4605 & 0.5855
& 0.4135 & 0.9170 \\

\hline
 & \multicolumn{5}{l}{$\lrmv=5$}
 & \multicolumn{5}{l}{$\lrmv=10$}\\

V, max ($\theta_{\rm f2}$,10000) & 0.9663 & 0.6138 & 0.9147 & 0.2776
& 0.9833* & 0.6631
& 0.5874 & 0.5407 & 0.2317 & 0.9339 \\

V, min ($\theta_{\rm f2}$,10000) & 0.9909* & 0.7515 & 0.9754? &
0.6519 & 0.9864* & 0.9728?
& 0.8389 & 0.8678 & 0.7187 & 0.9685 \\

\hline
\hline
\end{tabular}
\end{center}
\label{1p_tab_vk}
\end{table*}

\begin{table*}
\caption{One-point statistics as a function of the Galactic cut, $|b|=15^{\circ}$,
$20^{\circ}$, $25^{\circ}$ and $30^{\circ}$, imposed in addition to the KQ75 mask
for $\lrmv=1$ and $2$.
10000 simulations were performed for V band, with smoothing $\rm{FWHM}=43.485'$.}
\begin{center}
\begin{tabular}{r|lllll|lllll}

\hline
\hline

 & \multicolumn{5}{l}{$\lrmv=1$}
 & \multicolumn{5}{l}{$\lrmv=2$}\\

 $|b|$(degree)
 & NS & GN & GS & EN & ES
 & NS & GN & GS & EN & ES\\

\hline

 NUM(max) 15 & 0.5818  & 0.3038 & 0.7204 & 0.4550  & 0.6224 & 0.5521  & 0.3483 & 0.8076 & 0.3983  & 0.6242 \\
          20 & 0.4414  & 0.3606 & 0.5036 & 0.3480  & 0.5576 & 0.4110  & 0.4099 & 0.6079 & 0.2775  & 0.5359 \\
          25 & 0.4747  & 0.3329 & 0.5817 & 0.3205  & 0.5952 & 0.4183  & 0.3706 & 0.6329 & 0.2659  & 0.5143 \\
          30 & 0.3189  & 0.2562 & 0.4371 & 0.2720  & 0.3507 & 0.2365  & 0.2598 & 0.4745 & 0.2055  & 0.2871 \\

 \hline
 NUM(min) 15 & 0.9839* & 0.9672  & 0.8194 & 0.9849* & 0.9280 & 0.9827* & 0.9712  & 0.7839 & 0.9791* & 0.9127 \\
          20 & 0.9932* & 0.9902* & 0.8124 & 0.9935* & 0.9557 & 0.9938* & 0.9931* & 0.7843 & 0.9911* & 0.9361 \\
          25 & 0.9802* & 0.9812* & 0.7173 & 0.9832* & 0.9322 & 0.9855* & 0.9840* & 0.7055 & 0.9814* & 0.9264 \\
          30 & 0.9156  & 0.9917* & 0.3048 & 0.9731? & 0.7751 & 0.9379  & 0.9952* & 0.2802 & 0.9733? & 0.7995 \\

 \hline
MEAN(max) 15 & 0.1553 & 0.0507 & 0.3109 & 0.1326 & 0.2744 & 0.0666 & 0.5647 & 0.5054 & 0.1618 & 0.1535 \\
          20 & 0.3028 & 0.0924 & 0.3289 & 0.1662 & 0.4353 & 0.1026 & 0.7123 & 0.5231 & 0.1505 & 0.3231 \\
          25 & 0.4125 & 0.1132 & 0.4237 & 0.2434 & 0.4702 & 0.1462 & 0.5455 & 0.6947 & 0.2206 & 0.3787 \\
          30 & 0.4713 & 0.1369 & 0.4425 & 0.2914 & 0.4812 & 0.1585 & 0.3582 & 0.5935 & 0.2892 & 0.3596 \\

 \hline
MEAN(min) 15 & 0.4629 & 0.6680 & 0.3954 & 0.4417 & 0.4593 & 0.6154 & 0.0837 & 0.2139 & 0.4418 & 0.5819 \\
          20 & 0.4863 & 0.6534 & 0.5305 & 0.5805 & 0.4183 & 0.6997 & 0.0354 & 0.3214 & 0.5704 & 0.5198 \\
          25 & 0.4539 & 0.6290 & 0.4977 & 0.6005 & 0.3845 & 0.6756 & 0.0681 & 0.2671 & 0.5768 & 0.4469 \\
          30 & 0.4520 & 0.5258 & 0.6342 & 0.5708 & 0.4604 & 0.7103 & 0.1645 & 0.6419 & 0.5195 & 0.5961 \\

 \hline
 VAR(max) 15 & 0.0028* & \underline{0.0006}* & 0.4936 & \underline{0.0005}* & 0.3152 & 0.0460  & 0.0017* & 0.7935 & 0.0023* & 0.6481 \\
          20 & 0.0033* & 0.0014* & 0.4748 & \underline{0.0010}* & 0.3046 & 0.0493  & 0.0024* & 0.7675 & 0.0039* & 0.6652 \\
          25 & 0.0027* & 0.0028* & 0.3957 & 0.0028* & 0.1961 & 0.0409  & 0.0027* & 0.6755 & 0.0050* & 0.5817 \\
          30 & 0.0020* & 0.0042* & 0.2589 & \underline{0.0010}* & 0.1514 & 0.0195* & 0.0037* & 0.5766 & 0.0022* & 0.5319 \\
 \hline
 VAR(min) 15 & \underline{0.0013}* & \underline{0.0006}* & 0.3974 & \underline{0.0005}* & 0.2368 & 0.0183* & 0.0013* & 0.7396 & 0.0015* & 0.5043 \\
          20 & 0.0019* & \underline{0.0011}* & 0.4116 & \underline{0.0009}* & 0.2532 & 0.0235? & 0.0012* & 0.7340 & 0.0024* & 0.5714 \\
          25 & 0.0023* & 0.0020* & 0.3898 & 0.0032* & 0.1640 & 0.0207* & 0.0014* & 0.7382 & 0.0063* & 0.5098 \\
          30 & 0.0026* & 0.0029* & 0.3302 & 0.0020* & 0.1716 & 0.0176* & 0.0020* & 0.7049 & 0.0040* & 0.5517 \\

 \hline
KURT(max) 15 & 0.8411 & 0.8573 & 0.5413 & 0.7013 & 0.8641 & 0.8571 & 0.7929 & 0.7216 & 0.6022 & 0.9074 \\
          20 & 0.7587 & 0.8111 & 0.4698 & 0.6584 & 0.7888 & 0.7785 & 0.7396 & 0.6777 & 0.5720 & 0.8724 \\
          25 & 0.7332 & 0.8665 & 0.3666 & 0.6614 & 0.8423 & 0.7816 & 0.8404 & 0.7091 & 0.6413 & 0.9210 \\
          30 & 0.6148 & 0.8270 & 0.3273 & 0.5461 & 0.7880 & 0.6786 & 0.7711 & 0.7072 & 0.4947 & 0.8280 \\

 \hline
KURT(min) 15 & 0.8666 & 0.8631 & 0.6038 & 0.7742 & 0.8656 & 0.9641 & 0.8418 & 0.8050 & 0.7753 & 0.9310 \\
          20 & 0.8556 & 0.8654 & 0.5891 & 0.6777 & 0.9060 & 0.9526 & 0.8694 & 0.7891 & 0.7309 & 0.9439 \\
          25 & 0.7774 & 0.7677 & 0.5720 & 0.5745 & 0.8959 & 0.9048 & 0.6792 & 0.7995 & 0.6469 & 0.9033 \\
          30 & 0.8698 & 0.8064 & 0.7411 & 0.7471 & 0.9221 & 0.9510 & 0.7932 & 0.8945 & 0.8110 & 0.9146 \\

 \hline
 \hline
\end{tabular}
\end{center}
\label{tab_galcut_rl0102}
\end{table*}

\begin{table*}
\caption{Continued from Table \ref{tab_galcut_rl0102} with $\lrmv=5$
and $10$.}
\begin{center}
\begin{tabular}{r|lllll|lllll}
 \hline
 \hline
 & \multicolumn{5}{l}{$\lrmv=5$}
 & \multicolumn{5}{l}{$\lrmv=10$}\\

 $|b|$(degree)
 & NS & GN & GS & EN & ES
 & NS & GN & GS & EN & ES\\

 \hline
 NUM(max) 15 & 0.7432 & 0.2372 & 0.9537 & 0.4920 & 0.7149 & 0.7470 & 0.3177 & 0.9212 & 0.3593 & 0.9189 \\
          20 & 0.6206 & 0.3685 & 0.7804 & 0.3241 & 0.6494 & 0.6891 & 0.3648 & 0.8352 & 0.3061 & 0.8972 \\
          25 & 0.6084 & 0.3158 & 0.7905 & 0.2675 & 0.6004 & 0.6783 & 0.4069 & 0.8071 & 0.3005 & 0.8891 \\
          30 & 0.3821 & 0.2499 & 0.5695 & 0.1833 & 0.3841 & 0.3867 & 0.3080 & 0.5927 & 0.2512 & 0.6930 \\

 \hline
 NUM(min) 15 & 0.9378 & 0.9597  & 0.6160 & 0.9598  & 0.6016 & 0.8918 & 0.9318  & 0.7042 & 0.9262 & 0.7854 \\
          20 & 0.9674 & 0.9822* & 0.6840 & 0.9779* & 0.6868 & 0.9574 & 0.9796* & 0.7329 & 0.9462 & 0.8530 \\
          25 & 0.9498 & 0.9755? & 0.6276 & 0.9531  & 0.6514 & 0.9375 & 0.9682  & 0.6480 & 0.8907 & 0.8752 \\
          30 & 0.8915 & 0.9885* & 0.2275 & 0.9234  & 0.3993 & 0.8764 & 0.9882* & 0.3297 & 0.8783 & 0.6698 \\

 \hline
MEAN(max) 15 & 0.3757 & 0.5227 & 0.2834 & 0.2356 & 0.3585 & 0.4453 & 0.4188 & 0.4247 & 0.6438 & 0.2208 \\
          20 & 0.4934 & 0.6329 & 0.3103 & 0.2488 & 0.5865 & 0.4512 & 0.6615 & 0.2511 & 0.5414 & 0.3296 \\
          25 & 0.7283 & 0.6669 & 0.6404 & 0.5735 & 0.7172 & 0.6779 & 0.5447 & 0.5576 & 0.6904 & 0.5320 \\
          30 & 0.7607 & 0.6678 & 0.6307 & 0.7316 & 0.6048 & 0.6975 & 0.4704 & 0.5787 & 0.7336 & 0.4436 \\

 \hline
MEAN(min) 15 & 0.2167 & 0.1231 & 0.3206 & 0.2959 & 0.2545 & 0.1851 & 0.2985 & 0.3055 & 0.2505 & 0.2442 \\
          20 & 0.2078 & 0.0765 & 0.4253 & 0.3014 & 0.1706 & 0.1919 & 0.1318 & 0.5056 & 0.2762 & 0.1873 \\
          25 & 0.0705 & 0.0318 & 0.2508 & 0.1136 & 0.0786 & 0.0767 & 0.1317 & 0.2935 & 0.2037 & 0.0598 \\
          30 & 0.0898 & 0.0223* & 0.5859 & 0.0642 & 0.3915 & 0.1743 & 0.1343 & 0.6256 & 0.1912 & 0.3039 \\

 \hline
 VAR(max) 15 & 0.1011 & 0.0597 & 0.7851 & 0.0533 & 0.5263 & 0.1112 & 0.1893 & 0.3943 & 0.2894 & 0.4045 \\
          20 & 0.1124 & 0.0637 & 0.7398 & 0.0501 & 0.4753 & 0.0984 & 0.1415 & 0.3727 & 0.3446 & 0.3012 \\
          25 & 0.0822 & 0.0420 & 0.5815 & 0.0352 & 0.3717 & 0.0595 & 0.1208 & 0.2426 & 0.2619 & 0.1737 \\
          30 & 0.0588 & 0.0391 & 0.4834 & 0.0225? & 0.4489 & 0.0604 & 0.2150 & 0.1922 & 0.2931 & 0.2051 \\
 \hline
 VAR(min) 15 & 0.1020 & 0.0385 & 0.6171 & 0.0192* & 0.5127 & 0.0246? & 0.1240 & 0.1634 & 0.1131 & 0.2558 \\
          20 & 0.1183 & 0.0285 & 0.6540 & 0.0195* & 0.5278 & 0.0261? & 0.0606 & 0.2471 & 0.1367 & 0.2534 \\
          25 & 0.1547 & 0.0230? & 0.7443 & 0.0423 & 0.4741 & 0.0466 & 0.0667 & 0.3211 & 0.2351 & 0.1873 \\
          30 & 0.1752 & 0.0260? & 0.7560 & 0.0294 & 0.6663 & 0.1199 & 0.1905 & 0.4644 & 0.3255 & 0.3856 \\

\hline
KURT(max) 15 & 0.9673  & 0.6796 & 0.8800  & 0.2059 & 0.9888* & 0.6835  & 0.6629 & 0.4533 & 0.2227 & 0.9488 \\
          20 & 0.9756? & 0.6771 & 0.9253  & 0.2814 & 0.9913* & 0.6926  & 0.6320 & 0.5131 & 0.2714 & 0.9477 \\
          25 & 0.9882* & 0.8077 & 0.9453  & 0.3999 & 0.9950* & 0.7609  & 0.7060 & 0.5327 & 0.3122 & 0.9657 \\
          30 & 0.9495  & 0.4888 & 0.9208  & 0.3806 & 0.9672  & 0.5898  & 0.5610 & 0.4905 & 0.3529 & 0.9347 \\
\hline
KURT(min) 15 & 0.9948* & 0.7858 & 0.9848* & 0.6574 & 0.9907* & 0.9859* & 0.8352 & 0.9078 & 0.7469 & 0.9780* \\
          20 & 0.9951* & 0.8475 & 0.9819* & 0.5897 & 0.9956* & 0.9841* & 0.8354 & 0.8896 & 0.6954 & 0.9844* \\
          25 & 0.9859* & 0.7030 & 0.9786* & 0.5293 & 0.9904* & 0.9487  & 0.6758 & 0.8421 & 0.5825 & 0.9551 \\
          30 & 0.9927* & 0.7418 & 0.9887* & 0.6970 & 0.9885* & 0.8970  & 0.5917 & 0.7800 & 0.6058 & 0.8987 \\

\hline
\hline
\end{tabular}
\end{center}
\label{tab_galcut_rl0510}
\end{table*}

\subsection{Two-point results} \label{2p_results}
We study the correlation functions of hot and cold spots for the
V-band in five sky coverages with three kinds of temperature
threshold applied -- no threshold ($\rm T_{max}>-\infty$, $\rm
T_{min}<\infty$), $1\sigma_{\rm sky}$ ($\rm T_{max}>1\sigma_{\rm
sky}$ , $\rm T_{min}<-1\sigma_{\rm sky}$) and $2\sigma_{\rm sky}$
($\rm T_{max}>2\sigma_{\rm sky}$, $\rm T_{min}<-2\sigma_{\rm sky}$),
where $\sigma_{\rm sky}$ is the standard deviation of the
temperature of all the valid pixels in a certain sky region. For the
sky region outside the KQ75 mask, we compute correlation functions
from $0^\circ$ to $180^\circ$, divided into 1000 bins, and from
$0^\circ$ to $90^\circ$ with 500 bins for the four hemispheres in
question, since in these cases pairs with separations larger than
$90^\circ$ are so rare, in particular for the 2$\sigma_{\rm sky}$
threshold, that the correlation functions become strongly
oscillatory, affecting the accuracy of inversion of the covariance
matrix.

Large-scale modes do affect the confidence level of our one-point
statistics, especially the variance, the temperature-temperature
correlation of local extrema at zero lag. Thus, conceivably, such a
low-$\ell$ subtraction is also of importance to the correlation
study. According to the theoretical calculation of T-T correlation
by \citet{heavens_etal_1999}, there are oscillation features on
angular scales from 10 to 100 arcmin, however, we will not attempt
to investigate these features because of the observational
resolution and smoothing procedure during the data processing. We
expect stronger signal in this angular range by the forthcoming
higher resolution full-sky CMB survey, the \textit{PLANCK} mission.

\subsubsection{T-T correlation} \label{sec_tt}

\begin{figure*}
\begin{center}
\includegraphics[width=\textwidth,trim=0cm 4.0cm 0.7cm 0.5cm]{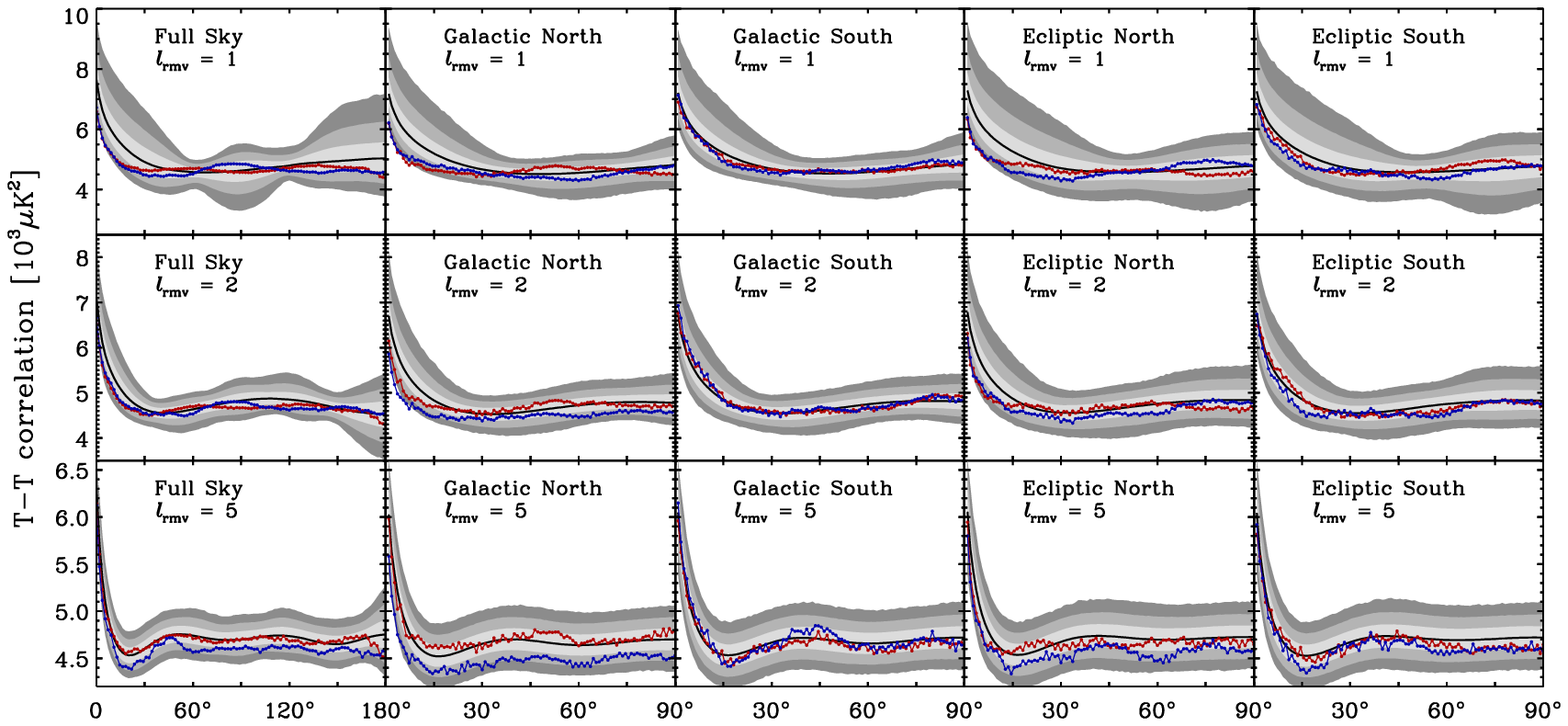}
\caption{Temperature-temperature correlation functions of hotspots
and cold spots without temperature threshold, for five sky-coverages
adopted in this paper. The red (blue) dots correspond to the
correlation function bins of hotspots (cold spots) observed in the
\WMAP\ V-band data. The light, middle and dark gray shaded bands
show, respectively, the 68.26\%, 95.44\% and 99.74\% confidence
regions determined from 10000 MC simulations, and the black solid
line shows the median. Since the statistical bands of hotspots and
cold spots are quite identical, we plot averaged bands here. The
correlation functions on both full sky and hemispheres are rebinned
to 100 bins.} \label{fig_TT_full}
\end{center}
\end{figure*}

We show the temperature-temperature correlation functions without
threshold applied in Figure \ref{fig_TT_full}.\footnote{The full
categories of correlation function plots, including T-T and P-P
correlations of all the cases discussed in this paper, can be
downloaded from this address:
http://www.mpa-garching.mpg.de/$\sim$houzhen/cf.tar.gz} The
correlation functions have been rebinned to 100 bins for both full
sky and hemispheres and the first bins are dropped.

\begin{figure}
\includegraphics[width=0.80\textwidth,trim=0 0 0 0.5cm]{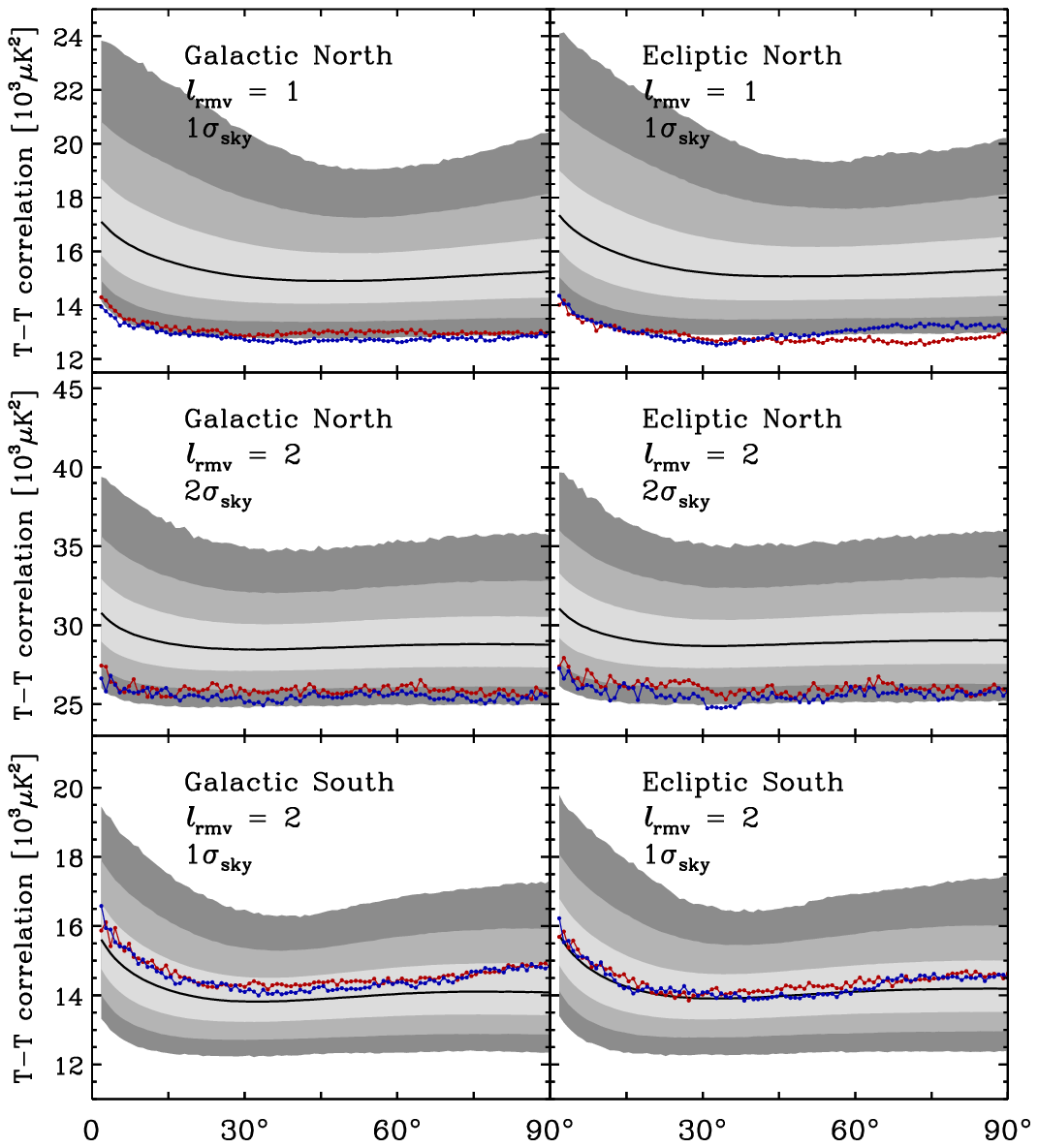}
\caption{Temperature-temperature correlation functions of hotspots
and cold spots valid for thresholds (1$\sigma_{\rm sky}$ and
$2\sigma_{\rm sky}$). The nomenclature of the dots and bands follows
the same style as Figure \ref{fig_TT_full}.}

\label{fig_TT}
\end{figure}

\citet{spergel_etal_2003} and \citet{copi_etal_2008} report a lack
of correlated signal compared to the $\Lambda$CDM model for angular
scales greater than $60^\circ$ for \WMAP1 and \WMAP5, respectively.
However, as shown by the top and middle panels of Figure
\ref{fig_TT}, there is no significant lack of correlation for the
hot and cold spots on corresponding scales, when computed on either
the full-sky or hemispheres. However, what is of note is the lack of
variance in the bins on these angular scales -- almost all lie
within the $2\sigma$ confidence region. In fact, the correlation
properties of the hot and cold spots show several interesting
properties as a function of angular scale and temperature threshold,
as we discuss below.

In the case of the full sky analysis with $\lrmv=1$ and without any
threshold applied, it should be apparent that there is a strong
suppression of the correlation function for the bins of both hot and
cold spots on scales less than about $20^\circ$ (corresponding to
the typical scale of $\ell=10$), to amplitudes around or even below
the lower limit of the Gaussian $3\sigma$ confidence region. Similar
behaviour is seen on scales less than $10^\circ$ in the Galactic and
Ecliptic north ($\lrmv=1$), whereas the southern sky shows quite a
good agreement with theoretical expectations. This corresponds to a
hemispherical asymmetry of power seen in the fluctuations of local
extrema. The suppression still exists after quadrupole subtraction,
in particular the correlation of cold spots in the Galactic north is
lower than the $3\sigma$ confidence region at small scales. After
removing the modes for $\lrmv=5$ and $10$, the small-scale
suppression become less significant and the hotspots correlation
shows perfect agreement with the median, especially on the full-sky,
which gives strong and self-consistent evidence that large-scale
moments ($\ell \le 10$) affect the small-scale
temperature-correlation of local extrema. Notice that the
distribution of correlations becomes less structured on large-scale
if higher moments are subtracted.

The angular structure of correlations for hot spots and cold spots
should be identical in theory, but the discrepancy between them gets
larger as higher moments are removed on both the full-sky and two
northern hemispheres. The hotspots are quite consistent with the
median of the predicted curves, whereas the cold spots on NS, GN and
EN are less correlated in the cases $\lrmv=5$ and 10.

The application of temperature thresholds to the set of hot and cold
spots included in the correlation study is also revealing (Figure
\ref{fig_TT}). There are strong suppressions on the full-sky and
northern hemispheres at all scales if the $1\sigma_{\rm sky}$ or
$2\sigma_{\rm sky}$ threshold is applied. For $\lrmv=1$ and 2 on GN
and EN, most bins are suppressed to the lower limit of $3\sigma$
confidence region, and many are apparently under the lower limit of
the $3\sigma$ confidence region defined by simulations. Conversely,
results for the southern sky still indicate agreement with the
best-fitting cosmological model. However, it should be recognised
that, apart from an overall scale-factor, the detailed shape of the
observed point-sets is consistent with the median at all other
scales. This implies that the T-T properties arise not from an
anomalous spatial clustering (or anti-clustering of local extrema),
but as a consequence of the statistical nature of the observed
temperature distribution. Specifically, it is conceivable that the
local extrema with higher temperatures in the northern sky are not
extreme enough.

Investigation of the corresponding cases for the P-P correlation function
should then also prove revealing.

We calculate the $\chi^2_s$ value defined by Eq.\ref{eq_chisqr}
based on the transformed distribution of the correlation functions
to quantify the degree of agreement between the observed data and
our Gaussian simulations. The fractions of simulations with
$\chi^2_s$ value lower than observed are listed in Table
\ref{tab_chisqr_tt}, with those larger than 0.975 underlined. All
the values after $\lrmv=10$ become normal. Notice that there are
cases with very low $p$-values, implying very, in some cases overly,
good agreement with the median. The disagreement of the small-scale
correlation without an applied threshold on full-sky coverage cannot
be distinguished by this full-scale analysis, and additional
small-scale analysis is necessary.

\begin{table*}
\caption{The frequencies of $\chi^2_s$ values from simulations with lower amplitude than
the \WMAP\ data as determined for the transformed T-T correlation functions. The values
higher than 0.9750 are underlined to demonstrate rejection of the Gaussian hypothesis at
$2\sigma$ level.}
\begin{center}
\begin{tabular}{c|lllll|lllll}

\hline \hline

 $\chi^2_{s}(TT)$-frequencies
 & \multicolumn{5}{l}{$\lrmv=1$}
 & \multicolumn{5}{l}{$\lrmv=2$}\\

 Thresholds
 & NS & GN & GS & EN & ES
 & NS & GN & GS & EN & ES\\

\hline

$\rm T_{max}>-\infty$ & 0.5774 & 0.9156 & 0.0077 & 0.6503 & 0.1141 &
0.5379 & 0.5070 & 0.0942 & 0.5199
& 0.0375 \\
$\rm T_{min}<\infty$ & 0.7595 & 0.7728 & 0.1413 & 0.7883 & 0.2150 &
0.4973 & 0.9488 & 0.0672 & 0.5863
& 0.2967 \\

$\rm T_{max}>1\sigma_{\rm sky}$ & \underline{0.9958} &
\underline{0.9947} & 0.1553 & \underline{0.9994} & 0.4059 & 0.9219 &
\underline{0.9968} & 0.4824 & \underline{0.9983}
& 0.2031 \\
$\rm T_{min}<-1\sigma_{\rm sky}$ & \underline{0.9947} &
\underline{0.9991} & 0.2315 & \underline{0.9973} & 0.3624 & 0.9470 &
\underline{0.9944} & 0.4083 & \underline{0.9892}
& 0.1161 \\

$\rm T_{max}>2\sigma_{\rm sky}$ & \underline{0.9919} &
\underline{0.9795} & 0.3493 & \underline{0.9955} & 0.4131 & 0.8506 &
0.9739 & 0.5942 & 0.9746
& 0.5306 \\
$\rm T_{min}<-2\sigma_{\rm sky}$ & \underline{0.9815} &
\underline{0.9928} & 0.0575 & \underline{0.9973} & 0.0627 & 0.7562 &
\underline{0.9887} & 0.6556 & \underline{0.9916}
& 0.6736 \\

 & \multicolumn{5}{l}{$\lrmv=5$}
 & \multicolumn{5}{l}{$\lrmv=10$}\\

$\rm T_{max}>-\infty$ & 0.0962 & 0.2958 & 0.1807 & 0.3249 & 0.2915 &
0.0233 & 0.0473 & 0.1605
& 0.1377 & 0.4012 \\
$\rm T_{min}<\infty$ & 0.8520 & 0.9400 & 0.2723 & 0.8133 & 0.5824 &
0.7731 & 0.7038 & 0.1791
& 0.6973 & 0.7379 \\

$\rm T_{max}>1\sigma_{\rm sky}$ & 0.8714 & 0.9628 & 0.2982 &
\underline{0.9862} & 0.0029 & 0.8287 & 0.7957 & 0.0259
& 0.6694 & 0.0186 \\
$\rm T_{min}<-1\sigma_{\rm sky}$ & 0.9394 & 0.8610 & 0.2107 & 0.9732
& 0.2949 & 0.9505 & 0.6467 & 0.4162
& 0.7179 & 0.5368 \\

$\rm T_{max}>2\sigma_{\rm sky}$ & 0.7214 & 0.7034 & 0.3321 & 0.9540
& 0.5583 & 0.7839 & 0.3833 & 0.5271
& 0.1774 & 0.4144 \\
$\rm T_{min}<-2\sigma_{\rm sky}$ & 0.7261 & 0.8316 & 0.4718 & 0.8683
& 0.1474 & 0.8029 & 0.7024 & 0.0438
& 0.7411 & 0.4075 \\

\hline \hline

\end{tabular}
\label{tab_chisqr_tt}
\end{center}
\end{table*}

\subsubsection{P-P correlation} \label{sec_pp}

\begin{figure}
\includegraphics[width=0.80\textwidth,trim=0 0 0 0.5cm]{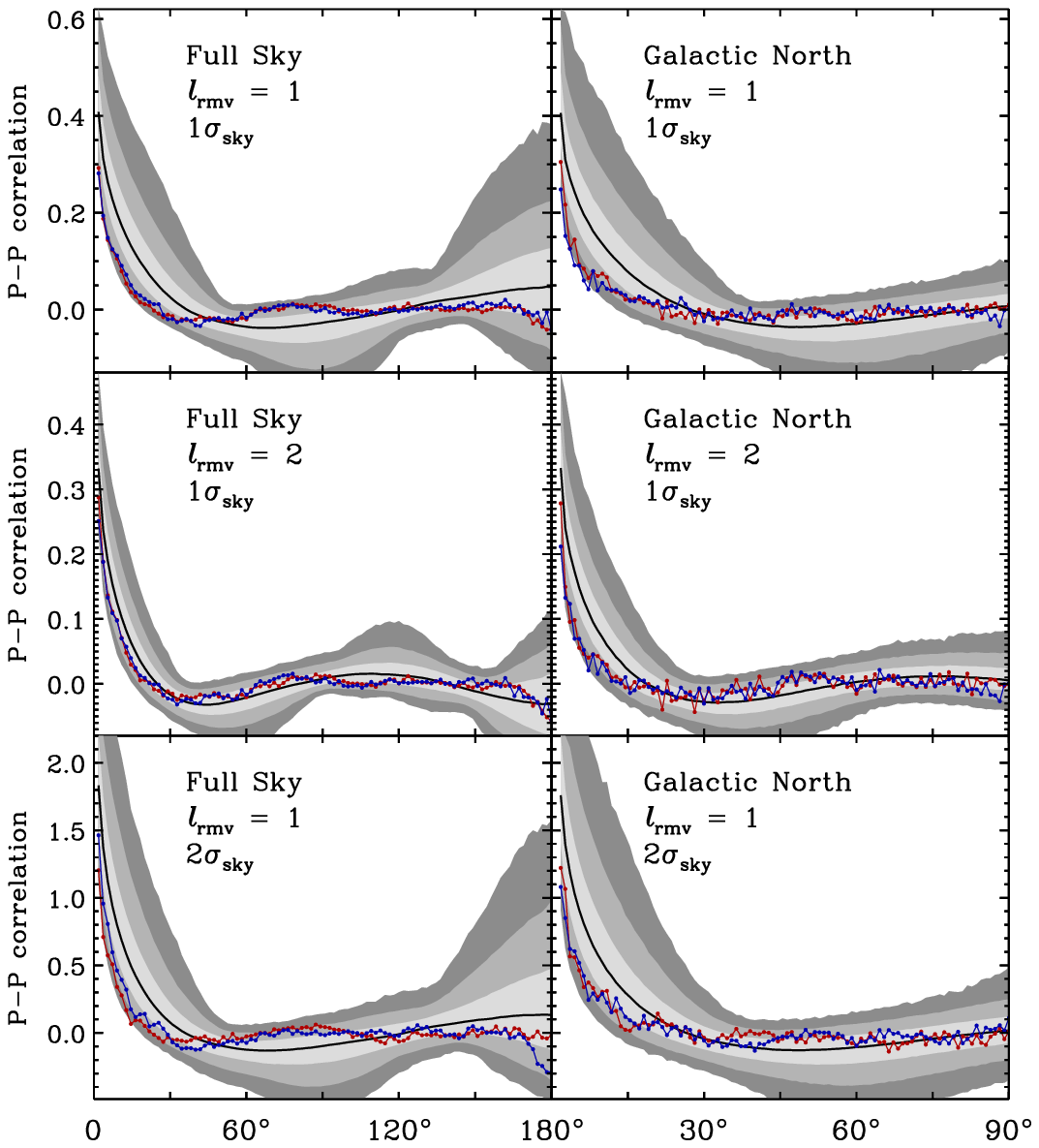}
\caption{Point-point correlation functions of hotspots and cold
spots valid for thresholds (1$\sigma_{\rm sky}$ and $2\sigma_{\rm
sky}$). 5000 simulations are performed. The nomenclature of the dots
and bands follows the same style as Figure \ref{fig_TT_full}.}
\label{fig_PP}
\end{figure}

The spatial clustering of local extrema quantified by the P-P
correlation function is analysed by comparing with 5000 simulations.
This analysis is also based on V-band data with a $43.485'$ FWHM
Gaussian beam smoothing applied. As with the temperature
correlation, we rebin the original structure to 100 bins for both
full sky and hemispheres and drop the first bins. Temperature
thresholds are also applied to select the valid spots, but only
their spatial positions are taken into account. Some of the results
are plotted in Figure \ref{fig_PP}. We take the simple averaged
confidence regions for hotspots and cold spots since the profiles of
these two are again essentially identical.

We do not show results corresponding to the application of no
threshold since the observed structure is trivial, essentially a
scatter around zero over all separation angles, and no
high-significance results are found under these conditions. This
means that the extrema in our data set are nearly-uniformly
distributed on the sphere. Instead, we focus on the correlation
structure of valid spots as a function of temperature thresholds. We
expect stronger clustering if we introduce a certain level threshold
that retains only higher amplitude peaks. We also introduce the
notion of a `clustering scale' by analogy with studies of the galaxy
two-point correlation function, The first decline to zero amplitude
defines  the characteristic radius of an extended hot or cold region
(or `cluster') and bumps on larger scales reflect the correlation
among `clusters'. Since the major clustering regions are determined
by large-scale modes, and the local extrema created by other modes
in these regions can be enhanced or suppressed by a certain level,
the features of the CMB local extrema correlation functions (both
T-T and P-P) represent the magnitude and scale of the hot and cold
regions of large-scale modes.

The results are consistent with our expectations. The
positive-valued structure demonstrates an approximate $30^\circ$
typical clustering scale before removing the quadrupole, and valid
points for the $2\sigma_{\rm sky}$ threshold show a much stronger
clustering amplitude (the left-three panels of Figure \ref{fig_PP}).
Moreover, the curves decline to zero faster after more moments are
removed, and are less structured on larger angular separations
because there is no sufficiently-long wavelength fluctuation to
correlate the local extrema on these scales. Thus, this typical
clustering scale corresponds to the spatial pattern of large-scale
modes.

There is a $3\sigma$-level suppression for both $1\sigma_{\rm sky}$
thresholded hot and cold spot correlations on scales less than
$20^\circ$ as computed for the full sky and $\lrmv=1$, and an
approximately $2\sigma$-level suppression for the $2\sigma_{\rm
sky}$ threshold, resulting from the less correlated northern sky.
After the first 5 or 10, moments are removed, better consistency
with the simulations is again seen yielding additional evidence that
the abnormal properties of the large angular-scale temperature
structure affects the statistical properties of our observed sky.

\begin{table*}
\caption{The frequencies of $\chi^2_s$ values from our simulations with amplitudes
lower than the \WMAP\ data for the transformed P-P correlation functions.}
\begin{center}
\begin{tabular}{c|lllll|lllll}

\hline \hline

 $\chi^2_{s}(PP)$-frequencies
 & \multicolumn{5}{l}{$\lrmv=1$}
 & \multicolumn{5}{l}{$\lrmv=2$}\\

 Thresholds
 & NS & GN & GS & EN & ES
 & NS & GN & GS & EN & ES\\

\hline

$\rm T_{max}>-\infty$ & 0.2896 & 0.8154 & 0.2632 & 0.5712 & 0.7857 &
0.1993 & 0.7161 & 0.2228 & 0.6309
& 0.7139 \\
$\rm T_{min}<\infty$ & 0.4936 & 0.2940 & 0.5508 & 0.0260 & 0.8302 &
0.3725 & 0.0458 & 0.3352 & 0.0055
& 0.6782 \\

$\rm T_{max}>1\sigma_{\rm sky}$ & 0.9139 & 0.7250 & 0.3310 & 0.8926
& 0.6794 & 0.4111 & 0.7817 & 0.5479 & 0.5460
& 0.4217 \\
$\rm T_{min}<-1\sigma_{\rm sky}$ & 0.8601 & 0.8479 & 0.1380 & 0.9190
& 0.5717 & 0.5655 & 0.8608 & 0.6618 & 0.3590
& 0.6761 \\

$\rm T_{max}>2\sigma_{\rm sky}$ & 0.9075 & 0.6543 & 0.4716 & 0.7523
& 0.7584 & 0.6277 & 0.6851 & 0.2454 & 0.5536
& 0.1664 \\
$\rm T_{min}<-2\sigma_{\rm sky}$ & 0.7554 & 0.5749 & 0.0364 & 0.8834
& 0.3955 & 0.0878 & 0.7157 & 0.7559 & 0.2125
& 0.6901 \\

 & \multicolumn{5}{l}{$\lrmv=5$}
 & \multicolumn{5}{l}{$\lrmv=10$}\\

$\rm T_{max}>-\infty$ & 0.4964 & 0.8852 & 0.0597 & 0.7042 & 0.7577 &
0.2212 & 0.9706 & 0.2468
& 0.6877 & 0.8827 \\
$\rm T_{min}<\infty$ & 0.3305 & 0.0296 & 0.1814 & 0.0083 & 0.7874 &
0.2064 & 0.1126 & 0.3476
& 0.0156 & 0.8086 \\

$\rm T_{max}>1\sigma_{\rm sky}$ & 0.4998 & 0.7692 & 0.3950 & 0.6966
& 0.0114 & 0.3355 & 0.9616 & 0.1350
& 0.7969 & 0.2687 \\
$\rm T_{min}<-1\sigma_{\rm sky}$ & 0.4470 & 0.6074 & 0.9161 & 0.7757
& 0.6521 & 0.5145 & 0.4639 & 0.6108
& 0.6913 & 0.7653 \\

$\rm T_{max}>2\sigma_{\rm sky}$ & 0.1050 & 0.2946 & 0.0795 & 0.5207
& 0.5148 & 0.1550 & 0.3517 & 0.0167
& 0.7053 & 0.1376 \\
$\rm T_{min}<-2\sigma_{\rm sky}$ & 0.2703 & 0.8831 & 0.7519 & 0.5813
& 0.0762 & 0.1392 & 0.4512 & 0.8849
& 0.5299 & 0.8398 \\

\hline \hline

\end{tabular}
\label{tab_chisqr_pp}

\end{center}
\end{table*}

Table \ref{tab_chisqr_pp} lists the frequency for which the
$\chi^2_s$ values computed for the simulations are lower than the
observed \WMAP\ values for the transformed P-P correlation
functions. No rejection is detected, supporting the consistency of
the spatial distribution of the observed local extrema with our
Gaussian simulations, over the whole angular range of our concern,
while suppression on small-scales is not revealed by this analysis.

\subsubsection{Analysis on small scales} \label{sec_2p_smsc}
Both the T-T (no threshold) and P-P (thresholded) correlation
functions record a $3\sigma$-level suppression on scales less than
$20^\circ$ for the full sky and $10^\circ$ for northern hemispheres.
\citet{tojeiro_etal_2006} find evidence for non-Gaussianity using
the P-P correlation function of local extrema in \WMAP1 data. The
correlation function they use is estimated and rebinned to 19
equally spaced bins up to a maximum separation of $30^\circ$. To
make a comparison with their results, we analyse the
$\chi^2_s$-frequency of rebinned T-T and P-P correlations on scales
less than $30^\circ$ with the first 17 bins for the full sky and on
scales less than $15^\circ$ with first 9 bins for the hemispheres.

\begin{table*}
\caption{$\chi^2_{s}$-frequency for T-T (no threshold) and P-P
(thresholds) correlations on small angular scales, up to the
maximum separation of $30^\circ$ for full sky and $15^\circ$ for
hemispheres.}
\begin{center}
\begin{tabular}{c|lllll|lllll}

\hline \hline

 $\chi^2_{s}$-fractions
 & \multicolumn{5}{l}{$\lrmv=1$}
 & \multicolumn{5}{l}{$\lrmv=2$}\\

 Thresholds
 & NS & GN & GS & EN & ES
 & NS & GN & GS & EN & ES\\

\hline

$\rm T^{TT}_{max}>-\infty$ & 0.9527 & \underline{0.9980} & 0.3425 &
\underline{0.9968} & 0.5120 & 0.8982 & 0.9682 & 0.3055 &
\underline{0.9828}
& 0.2511 \\
$\rm T^{TT}_{min}<\infty$ & 0.9682 & \underline{0.9968} & 0.0788 &
\underline{0.9936} & 0.6851 & 0.8026 & \underline{0.9994} & 0.5489 &
0.9674
& 0.3035 \\

$\rm T^{PP}_{max}>1\sigma_{\rm sky}$ & \underline{0.9851} &
\underline{0.9806} & 0.0384 & \underline{0.9944} & 0.4503 & 0.7419 &
\underline{0.9851} & 0.6912 & 0.9657
& 0.3813 \\
$\rm T^{PP}_{min}<-1\sigma_{\rm sky}$ & 0.9643 & \underline{0.9986}
& 0.1756 & \underline{0.9825} & 0.4230 & 0.7369 & \underline{0.9946}
& 0.7110 & 0.9570
& 0.2748 \\

$\rm T^{PP}_{max}>2\sigma_{\rm sky}$ & \underline{0.9850} & 0.9351 &
0.6916 & \underline{0.9894} & 0.6918 & 0.7164 & 0.9696 & 0.5238 &
0.9599
& 0.2122 \\
$\rm T^{PP}_{min}<-2\sigma_{\rm sky}$ & 0.8410 & 0.9509 & 0.0518 &
\underline{0.9768} & 0.2010 & 0.3378 & 0.9702 & 0.5541 & 0.8320
& 0.2622 \\

 & \multicolumn{5}{l}{$\lrmv=5$}
 & \multicolumn{5}{l}{$\lrmv=10$}\\

$\rm T^{TT}_{max}>-\infty$ & 0.1885 & 0.3167 & 0.0529 & 0.8406 &
0.2342 & 0.2272 & 0.4377 & 0.4106
& 0.0645 & 0.5216 \\
$\rm T^{TT}_{min}<\infty$ & 0.9002 & \underline{0.9934} & 0.2804 &
0.9226 & 0.2513 & 0.8145 & 0.8266 & 0.3016
& 0.7282 & 0.7371 \\

$\rm T^{PP}_{max}>1\sigma_{\rm sky}$ & 0.0770 & 0.7088 & 0.4147 &
0.9670 & 0.0196 & 0.5500 & 0.7012 & 0.3371
& 0.4481 & 0.3506 \\
$\rm T^{PP}_{min}<-1\sigma_{\rm sky}$ & 0.3565 & 0.8849 & 0.0636 &
0.7811 & 0.1294 & 0.8249 & 0.8276 & 0.2964
& 0.6549 & 0.3212 \\

$\rm T^{PP}_{max}>2\sigma_{\rm sky}$ & 0.2480 & 0.6266 & 0.0102 &
0.8651 & 0.0057 & 0.7978 & 0.6273 & 0.2420
& 0.2128 & 0.0298 \\
$\rm T^{PP}_{min}<-2\sigma_{\rm sky}$ & 0.3898 & 0.9387 & 0.2267 &
0.7418 & 0.8274 & 0.1856 & 0.0926 & 0.9066
& 0.0228 & 0.5863 \\

\hline \hline

\end{tabular}
\label{tab_chisqr_smsc}

\end{center}
\end{table*}

The $\chi^2_s$-frequencies are listed in Table
\ref{tab_chisqr_smsc}. High significance is still detected on the
northern hemispheres for $\lrmv=1$ and 2, which is consistent with
the conclusions of \citet{tojeiro_etal_2006} for the first year of
\WMAP\ data. However, the simulation sample volume of their work is
only 250 simulations. As presumed in Section \ref{sec_pp}, the
suppression of the P-P correlation is connected with the less
structured large-scale temperature distribution. However, as shown
in Table \ref{tab_chisqr_smsc} and Table \ref{tab_chisqr_pp}, this
suppression disappears after removing first 5 moments after which
good consistency is found for the whole angular range. Therefore, it
appears that the CMB moments from $\ell=2$ (monopole and dipole are
always subtracted in this work) to 5 affect the observed P-P
correlation structure. In the following section we attempt to
discuss how the specific nature of these modes can affect the
structure of the correlation functions and even the one-point
statistics.

\subsubsection{Correlation structure analysis}

\begin{figure}
\includegraphics[width=0.86\textwidth,trim=0 0 0 0.5cm]{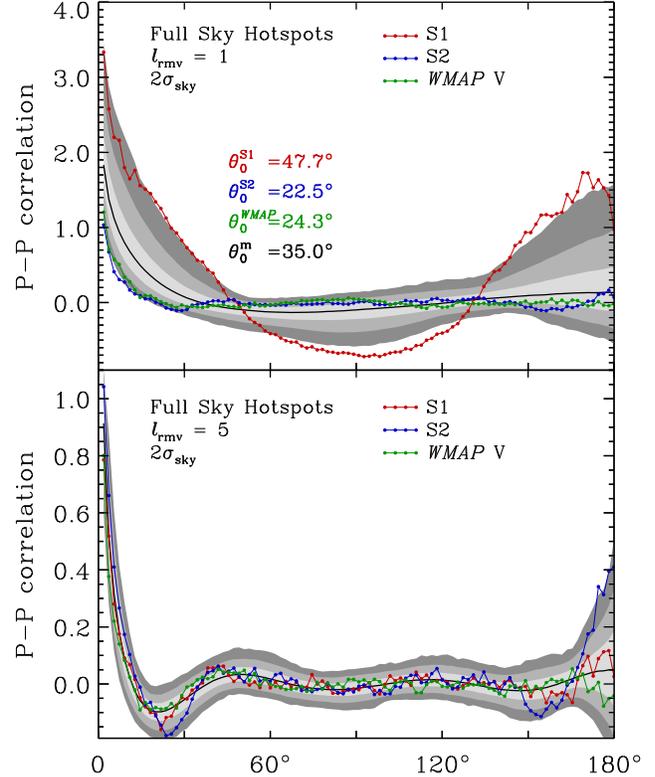}
\caption{The P-P hotspot correlation function on the full-sky with
$\lrmv=1$ and 5. The green dots corresponds to the \WMAP\ V band
data, and the red and blue dots are for two simulations marked S1
and S2, respectively. For $\lrmv=1$, $\theta_0$s give the angular
scales of the correlation functions at the first zero-crossing. The
correlations for $\lrmv=5$ are less structured and fit the simulated
statistical distribution, with few points outside $3\sigma$
confidence.} \label{fig_PP_structure}
\end{figure}

Since the correlation results, as well as our 1-point statistics,
for the observed sky show sensitivity to the first 5 large
angular-scale modes, it is worth analysing the large-scale phase
pattern with the help of the structure of the correlation functions.
The best-fitting multipoles as removed from the data are evaluated
on the sky region outside the KQ75 mask. It is inevitable that the
derived modes are coupled for such incomplete sky-coverage and our
best-fitting amplitudes also vary with the number of modes included
in the fit. However, such behaviour will also be found in the
reference set of simulations, whereas other unusual properties of
these modes, such as intrinsic correlations and alignments will not
be.  The analysis presented here will give a first assessment of the
relation between the large-scale temperature distribution and the
local extrema correlation structures.

We focus on the full-sky hotspots correlations, and in particular
the $2\sigma_{\rm sky}$ threshold for P-P, with only the monopole
and dipole removed.  An identical analysis can be applied to cold
spots. Two extremely behaved simulations, S1 and S2 hereafter, are
selected from our samples to make a comparison with the \WMAP\ one
in Figure \ref{fig_PP_structure}.  S1 corresponds to a
$\chi^2_s$-frequency 1.0000 and S2 to $\chi^2_s(30^\circ)$-frequency
0.9970.  S1 demonstrates strong spatial correlation over a range of
angular serparations before declining to zero at
$\theta^{S1}_0=47.7^\circ$ (within $2\sigma$ confidence).
Conversely, S2 is highly suppressed on small separation scales with
the zero-crossing at $\theta^{S2}_0=22.5^\circ$, and shows a similar
structure to the \WMAP\ data. The median determined from our
ensemble of simulations first crosses zero at $35^\circ$, with the
\WMAP\ one at $24.3^\circ$ around the $2\sigma$ lower bound.

\begin{figure*}
\begin{center}
\includegraphics[angle=90,width=0.48\textwidth]{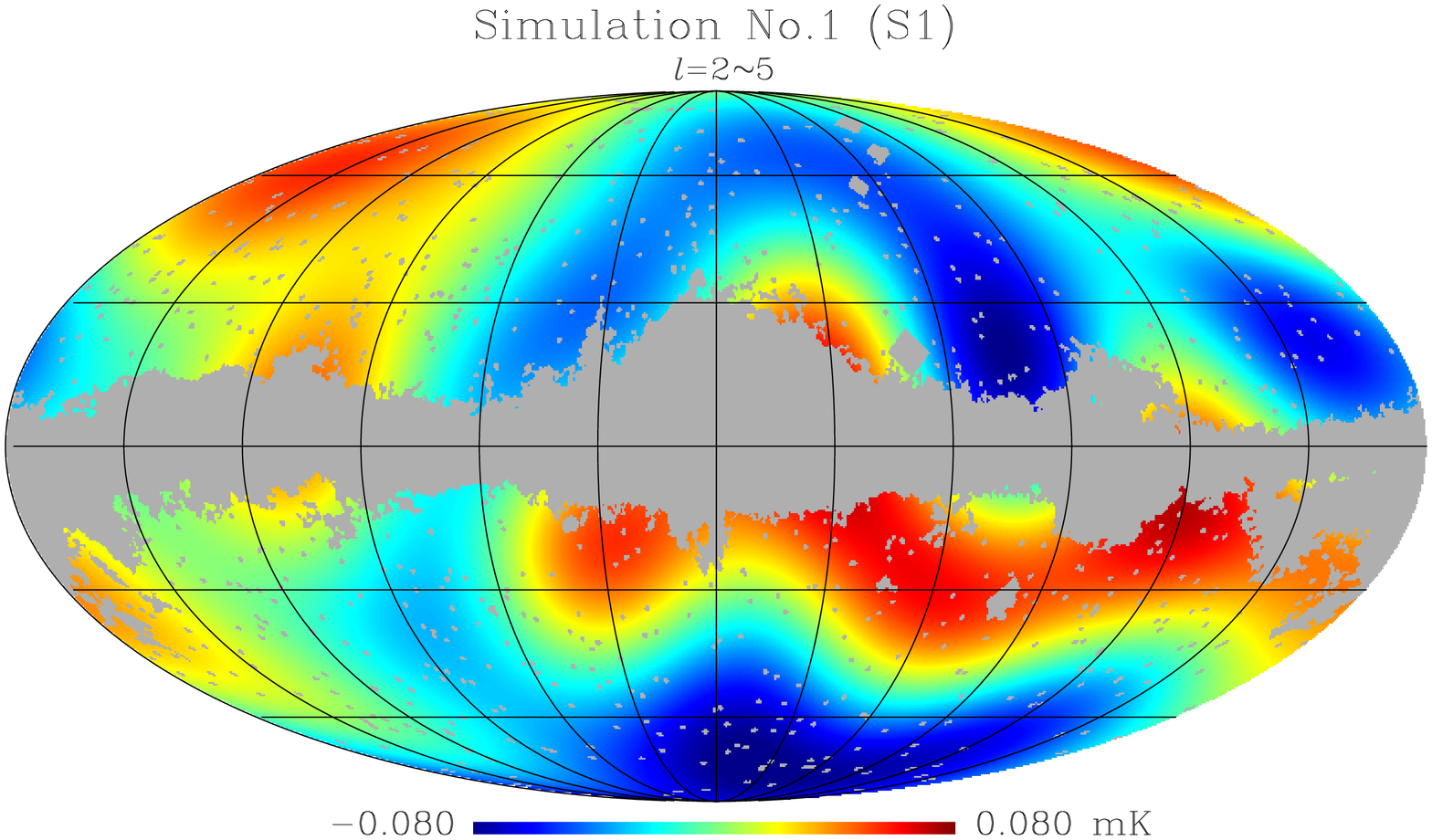}
\includegraphics[angle=90,width=0.48\textwidth]{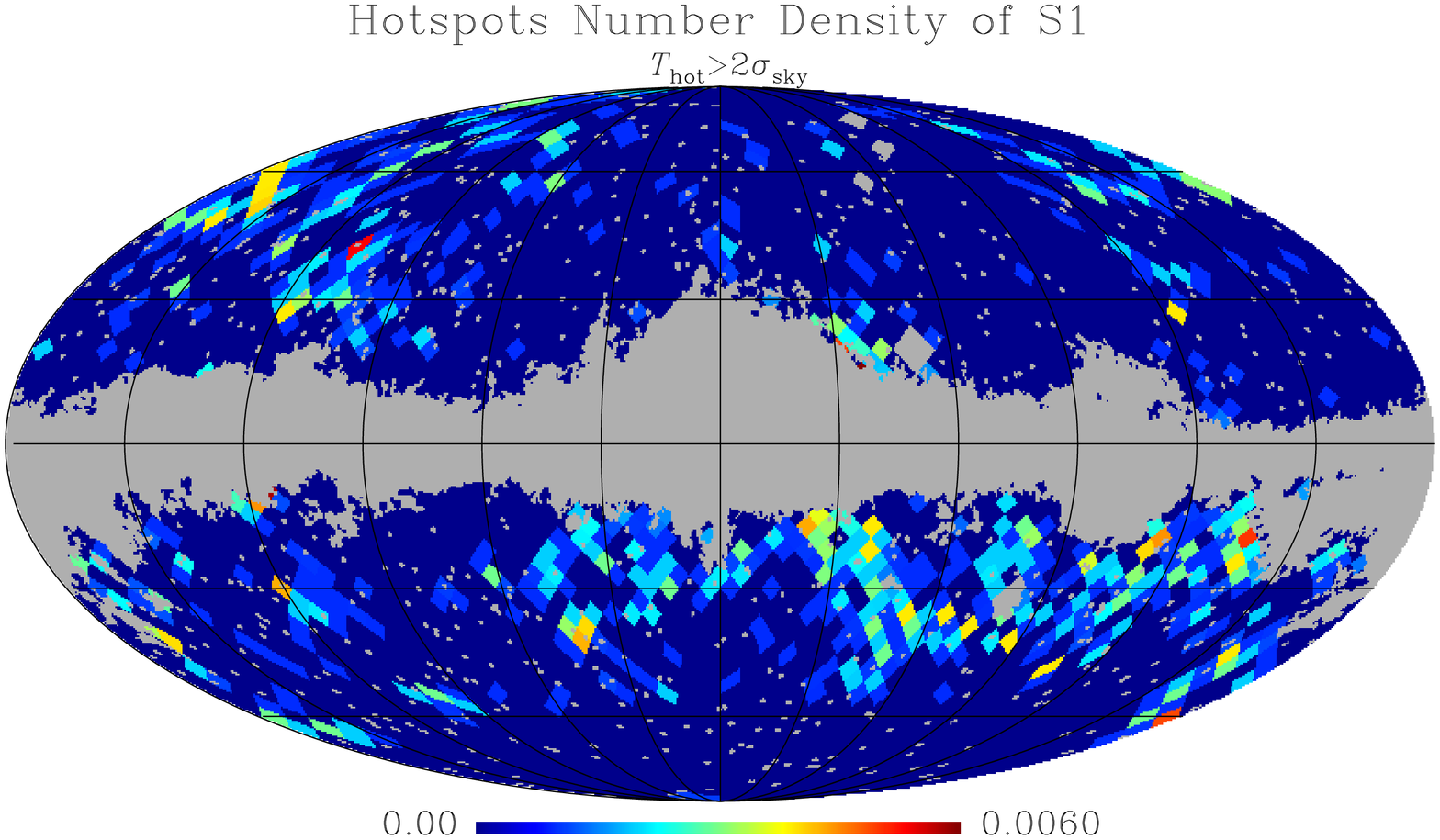}
\includegraphics[angle=90,width=0.48\textwidth]{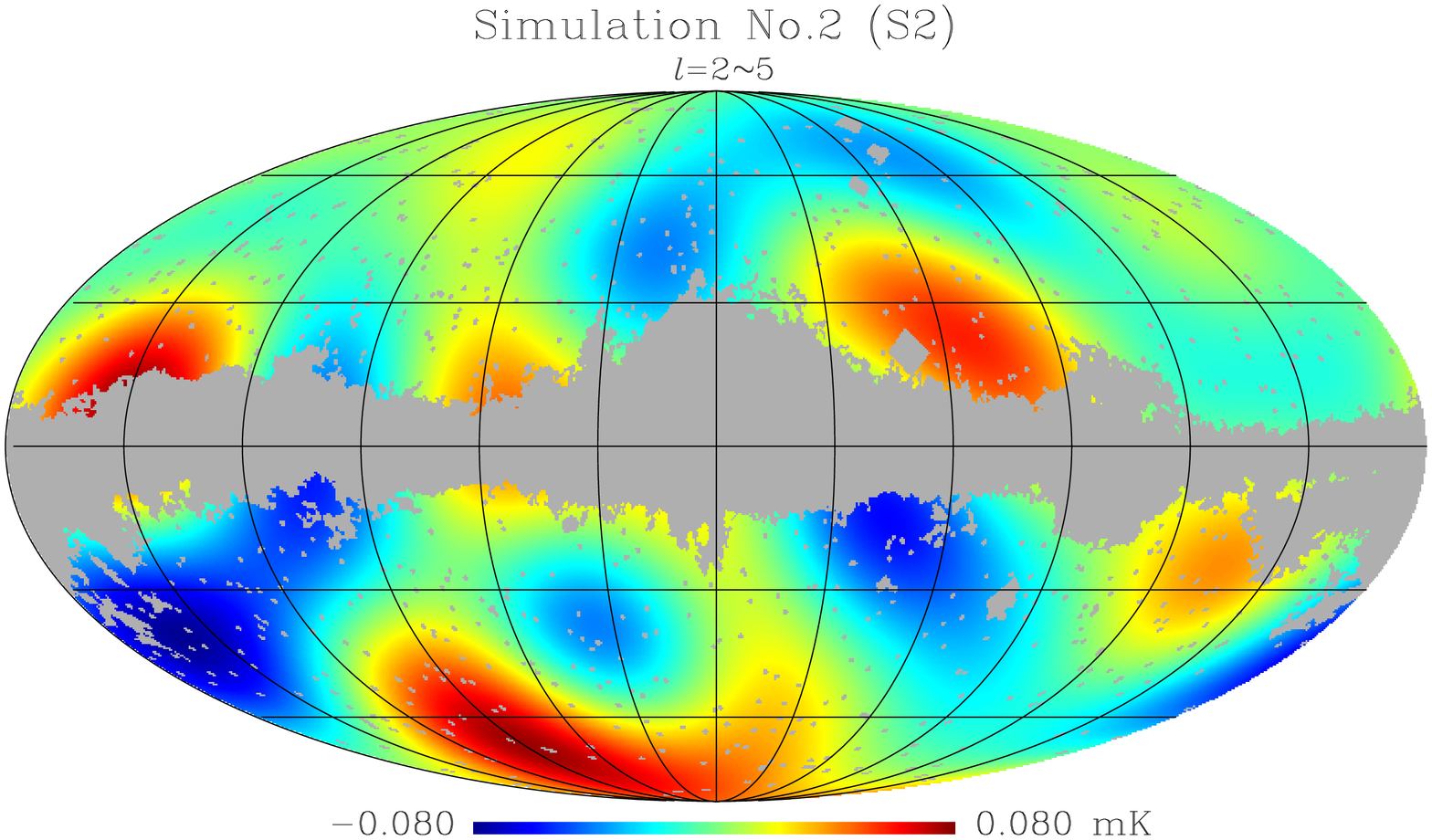}
\includegraphics[angle=90,width=0.48\textwidth]{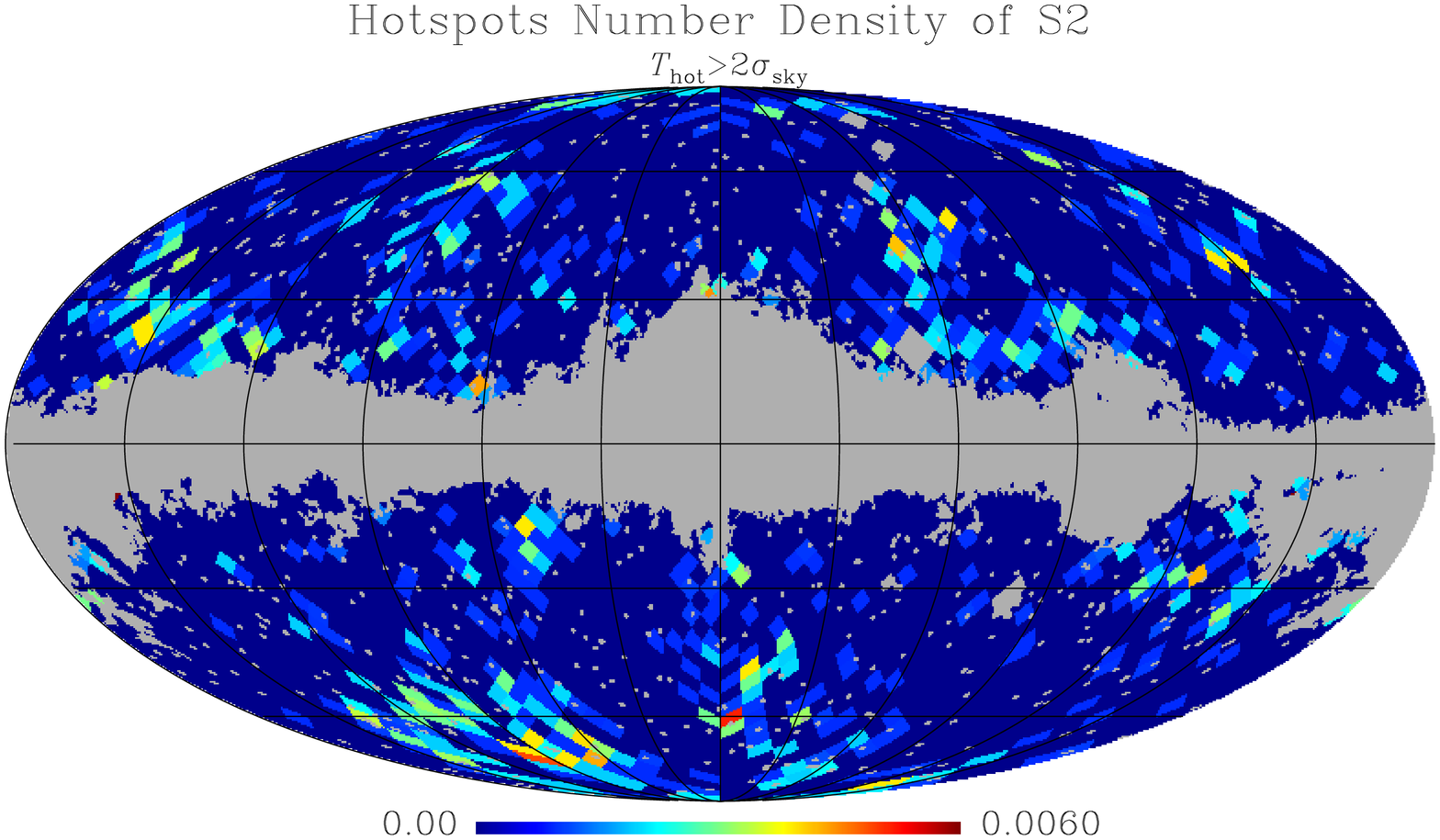}
\includegraphics[angle=90,width=0.48\textwidth]{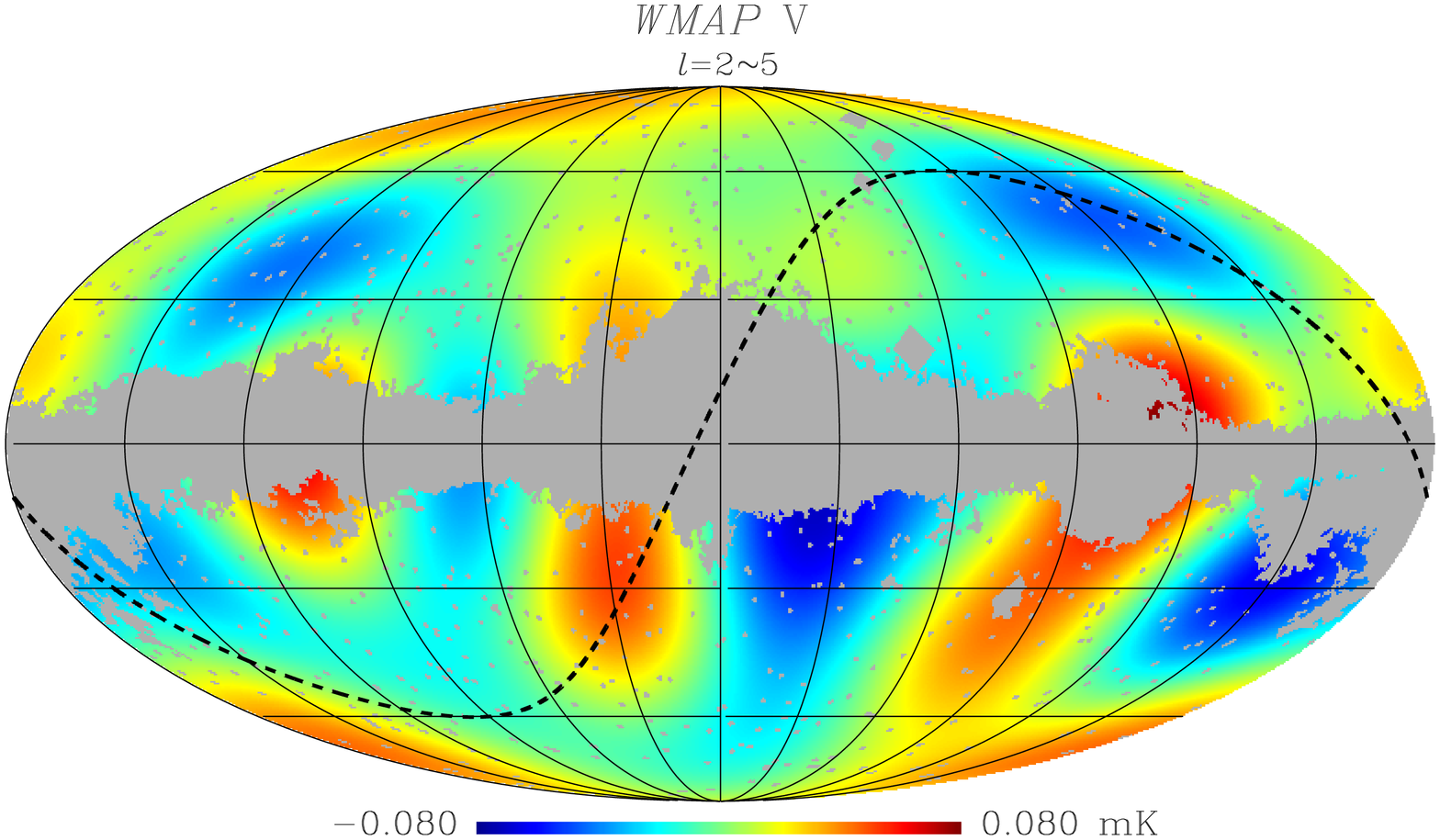}
\includegraphics[angle=90,width=0.48\textwidth]{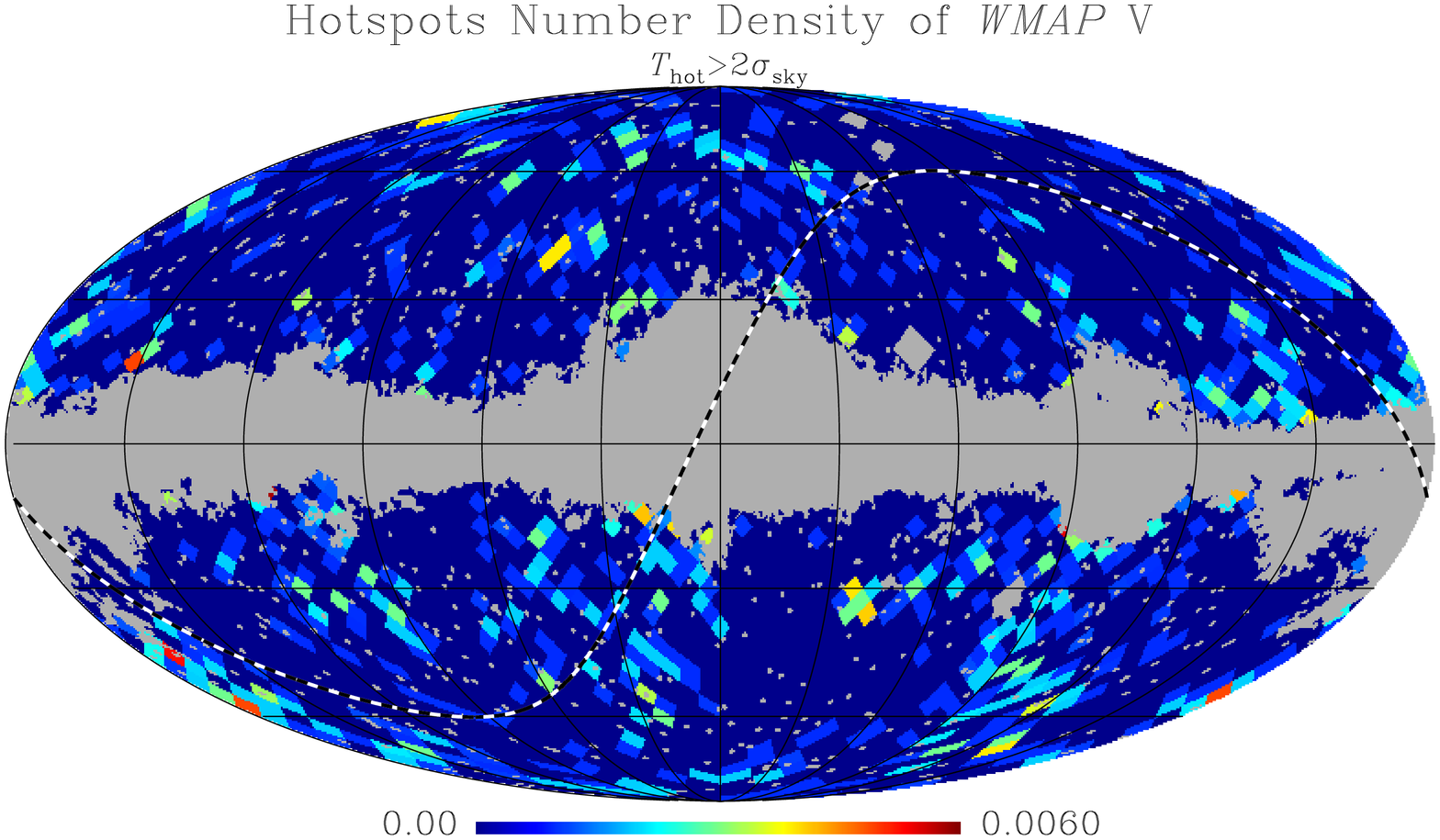}

\caption{{\it Left:} The coadded low order moments $\ell=2\sim5$
determined from (top to bottom) S1, S2 and the \WMAP\ V band, shown
in the Galactic coordinate system with a $30^\circ$ scalar
graticule. {\it Right:} The corresponding number-density of hotspots
for a $2\sigma_{\rm sky}$ temperature evaluated at resolution
$N_{\rm side}=16$. The KQ75 mask has been applied to the maps ad
shown in gray. The black dashed line on the bottom panel marks the
Ecliptic plane.} \label{fig_2l5}
\end{center}
\end{figure*}

We plot the coadded large-scale moments from $\ell=2$ to 5 in Figure
\ref{fig_2l5}, as well as the number-density of hotspots above
$2\sigma_{\rm sky}$, making a comparison with the correlation
structures. In particular, we note that the strong P-P correlation
of S1 below $\theta_0$ is related to the large and continuous hot
area, while the hot regions of S2 and \WMAP\ are smaller and more
scattered, which can be visually recognized from the number-density
realizations and gives strong connection between the structure of
extreme-hotspots and the large-scale modes of temperature
distribution. Therefore, the structures of the P-P correlation
function for $\lrmv=1$ are sensitive to the large-scale pattern of
temperature fluctuations. When modes corresponding to $\lrmv=5$ are
removed (Figure \ref{fig_PP_structure}), the correlation profile of
S1, S2 and the \WMAP\ data become less structured and more
consistent with the statistical distribution determined from the
simulations -- the points fit the median curve well with few points
outside $3\sigma$ confidence.

It is also notable from Figure \ref{fig_2l5} that a pronounced
north-south (both Galactic and Ecliptic) asymmetry is visible for
the hot and cold regions of the integrated large angular-scale
temperature distribution in the \WMAP\ data. This asymmetry is
consistent with the statistical behaviour of the temperature extrema
on the sky.

\section{CONCLUSIONS} \label{conclusion}
In this paper, we have investigated the statistical properties of local
extrema in the five-year \WMAP\ data release, and compared with
Gaussian simulations to determine whether the observed
universe is consistent with such processes. The analysis is carried out on
different frequency bands and five sky-coverages. Both the one-point
distribution and two-point correlation of local extrema have been
studied, including their dependence on large-scale CMB moments.

The hypothesis test on one-point statistics shows good consistency
of the number, mean, skewness and kurtosis values with the Gaussian
model used for comparison in most conditions. A few cases of
rejection occur for the mean value of cold spots, somewhat
consistent with the conclusion of \citet{LW04}, but these results
are not significant for all bands.  \citet{LW05} also found less
significant results for the mean values after applying smoothing to
the data, and attributed the earlier anomalies to the noise
properties of the first-year \WMAP\ data.

Our main results are the determination of anomalously low-variance
for the local extrema, and the presence of a north-south asymmetry,
the latter of which has been found to be indicated in other studies
of the \WMAP\ data using different statistical estimators. We find
that the data is inconsistent with simulations at the 95\% C.L. for
almost all frequency bands and both full-sky and the northern
hemispheres of two particular coordinate systems. We also find some
measurements on various scales/thresholds that are even lower than
the $3\sigma$ confidence region.

Our results argue against a residual-Galactic-foreground explanation
finding that the application of larger Galactic cuts, including
equatorial bands out to $30^\circ$ Galactic latitude, yields no
improvement in the consistency of data and Gaussian model
predictions. However, an improvement is observed after removing the
first 5 multipole moments, with the exception of the kurtosis values
for cold spots in the southern Ecliptic sky, and all the statistics
are well fitted after subtraction of the first 10 moments. This
strongly suggests that the anomalies are related to the first 5
large-scale moments, possibly extending to $\ell=10$. Further
confirmation is found via the two-point analysis.

Two kinds of two-point correlation analysis are performed to study
both the temperature behavior and spatial distribution of local
extrema. Limitations of the \WMAP\ angular resolution and data
processing imply that correlation features \citep{heavens_etal_1999}
on fine-scales ($10\sim100$ arcmin) cannot be investigated here. We
find that the T-T correlation functions with applied temperature
thresholds are dramatically suppressed on the full-sky and northern
hemispheres. The P-P correlations determined without any applied
threshold trivially oscillate around zero and lie well within the
confidence regions defined by simulations. However, both the T-T
correlation function without threshold and the P-P function with
thresholds are suppressed on small angular separations. The
$\chi^2_s$ values quantify the suppression level and some $3\sigma$
detections are found. All of this anomalous behaviour is improved
for $\lrmv=5$ and totally disappears for $\lrmv=10$.

Using two extreme simulations from our ensemble, we demonstrate a
connection between the P-P correlation structures and the pattern of
the corresponding fitted large-scale moments ($\ell=$ 2--5). The
number-density distribution of extreme-hotspots ($T>2\sigma_{\rm
sky}$) shows apparent correlation with hot regions of such
large-scale moments and so does the coldspots. For the \WMAP\ data,
it is also apparent that the northern hemisphere in both the
Galactic and ecliptic coordinate systems exhibit suppressed total
temperature fluctuations, which directly results in an insufficient
enhancement (suppression) of hot (cold) spots from moments $\ell >
5$. Therefore, the low-variance of local extrema is connected to
features of large-scale moments, as is the full-scale suppression of
the T-T correlation with thresholds.

\section*{ACKNOWLEDGEMENTS}
Z.H. acknowledges the support by Max-Planck-Gesellschaft Chinese
Academy of Sciences Joint Doctoral Promotion Programme
(MPG-CAS-DPP). We also thank Benjamin D. Wandelt, Hans K. Eriksen
and Cheng Li for useful discussions. Some of the results in this
paper have been derived using the HEALPix \citep{gorski_etal_2005}
software and analysis package. We acknowledge use of the Legacy
Archive for Microwave Background Data Analysis (LAMBDA).

\label{lastpage}
\end{document}